# Excitation and reception of magnetostatic surface spin waves in thin conducting ferromagnetic films by coplanar microwave antennas. Part II: Experiment

Charles Weiss[1], Matías Grassi[2], Yves Roussigné[3], Andrey Stashkevich[3], Thomas Schefer[1], Jerome Robert[2], Matthieu Bailleul[2] and Mikhail Kostylev[1*]

1. Department of Physics and Astrophysics, The University of Western Australia, Crawley, WA 6009, Australia
2. Institut de Physique et Chimie des Matériaux de Strasbourg, 23 rue du Loess, BP 43, 67034 Strasbourg Cedex 2, France
3. Laboratoire des Sciences des Procédés et des Matériaux, CNRS-UPR 3407, Université Sorbonne Paris Nord, Villetaneuse, France

Abstract

We report on propagating spin-wave spectroscopy measurements carried out on coplanar nano-antenna devices made from a Si/SiO$_2$/Ru(5nm)/Co(20)/Pt(5nm) film. The measurements were analyzed in detail by employing newly developed theoretical modeling and de-embedding procedures. The magnetic parameters of the film were determined by complementary Brillouin light scattering and ferromagnetic resonance measurements. The propagating spin wave signals could be accounted for quantitatively for the range of externally applied magnetic fields investigated in this study: 130-1500 Oe.

## 1. Introduction

The field of magnonics, which focuses on the study of standing and propagating spin waves, is of considerable research interest for potential microwave signal processing devices [1], sensing applications [2,3], traveling-spin-wave based logic [4], applications in quantum computing and information [5–7], as well as for neuromorphic reservoir computing using traveling spin waves [8].

Of these, traveling spin waves have additionally been useful for the characterization of magnetic materials. It has been suggested that measuring the spin-wave dispersion of ferromagnetic thin films allows for the direct extraction of the interface Dzyaloshinskii-Moriya interaction (iDMI) [9–16]. A well-known approach to probing traveling spin waves in a ferromagnetic film is through a spin wave delay line structure (spin-wave device) which employs two nano-scale antennas to excite and detect the spin waves [17–19]. In the first paper of this joint paper submission [20] we developed a self-consistent theoretical model for the excitation and reception of magnetostatic surface waves in thin strips of conductive ferromagnetic material by a pair of coplanar nano-antennas, and the model is implemented numerically.

The aim of this second experimental part was:

To investigate the typical level of losses in transmission and reflection and transmission and reflection bands of a microscopic spin wave device, employing a basic (i.e. not meander-like) coplanar line (CPL) antennas.

Additionally we wanted to compare the theory from part I with experimental measurements. However, to do this we first needed to understand how to de-embed the raw experimental data, in order to be able to carry out comparison of the experiment with the theory. Once the de-embedding of the experimental data had been completed, we wished to investigate whether the experimentally measured frequency dependences of the scattering parameters could be accurately fit with the theory developed in part I and whether we could use those fits in order to extract magnetic parameters of the ferromagnetic films.

---

* Corresponding author, mikhail.kostylev@uwa.edu.au

Furthermore, we wished to understand whether one could use the fits in order to extract values of device parameters that are difficult to model theoretically. For instance, losses in the feeding lines, level of off-band inductive coupling of the device input port to the output one and identify physical reasons for the coupling.

To perform the experimental measurements, a //Ru(5)/Co(20)/Pt(5) (where numbers in parenthesis are in nanometers) thin film sample was deposited onto a thermally oxidized intrinsic silicon substrate, and a ferromagnetic strip as well as a pair of coplanar nano-antennas was fabricated by lithographic techniques. The transmission characteristics of the spin-wave device were characterized using a vector network analyzer (VNA) and were compared to the numerically simulated transmission characteristics of the theoretical model. Due to the large parameter space of the spin-wave device parameters, Brillouin light scattering (BLS) and ferromagnetic resonance (FMR) measurements were employed in order to ease the determination of the device parameters.

## 2. Fabrication

The //Ru(5)/Co(20)/Pt(5) (where numbers in parenthesis are in nanometers) trilayer film was deposited on Si/SiO$_2$ substrates via DC magnetron sputtering using Ar as the sputtering gas at a base pressure of $2 \times 10^{-7}$ mbar. By means of laser beam lithography and ion beam etching the continuous film was patterned into a strip geometry with a strip width of 20 $\mu$m (*z*-direction Figure 1(a)) and a strip length of 80 $\mu$m (*x*-direction Figure 1(a)). On top of the strip we deposited ~80 nm of SiO$_2$ as a dielectric spacer layer. Next, a set of Ti(10)/Au(60) contact pads were deposited by means of electron beam evaporation and a laser beam lithography and lift-off process was employed to pattern the pads. Each set of pads (one for the excitation antenna and one for the receiving antenna) consisted of one central 'signal' pad and two outer 'ground' pads. These pads acted as an intermediate connector allowing for a connection between external circuit (via a set of microwave probes) and the CPL antennas themselves. Lastly a set of Ti(10)/Al(90) (CPL) antennas was fabricated on top of the strip using electron beam lithography, lift-off, and electron beam evaporation deposition. The antennas were fabricated in the Damon-Eshbach geometry and an example of one of the devices is displayed in Figure 1. Notice that the individual strips of the antennas start by overlapping with the gold contact pads and then narrow down to the CPL antenna sitting above the ferromagnetic strip. For the sample we fabricated a set of CPL antennas using the fabrication approach above. The following is the nominal antenna geometry: $w_g$ = 324 nm, $w$ = 648 nm, and $\Delta_g$ = 334 nm, where $w_g$ is the width of one of the ground conductors, $w$ is the width of the signal conductor, and $\Delta_g$ is the signal-ground conductor separation. Note that during the electron exposure of the photoresist in the lithographic process it is common for the electrons to scatter inside the photoresist resulting in a larger exposed area than what is nominally defined via the lithographic mask. As a result, the fabricated features tend to be larger than the nominal features. We estimated the true conductor widths from a scanning electron microscopy (SEM) image, shown in Figure 1(e), which corresponds to a device which had the same nominal geometry as the device investigated in this study. During the simulations we employed the measured antenna geometries, as we expected these to be closer to the true antenna geometry of the physical devices.

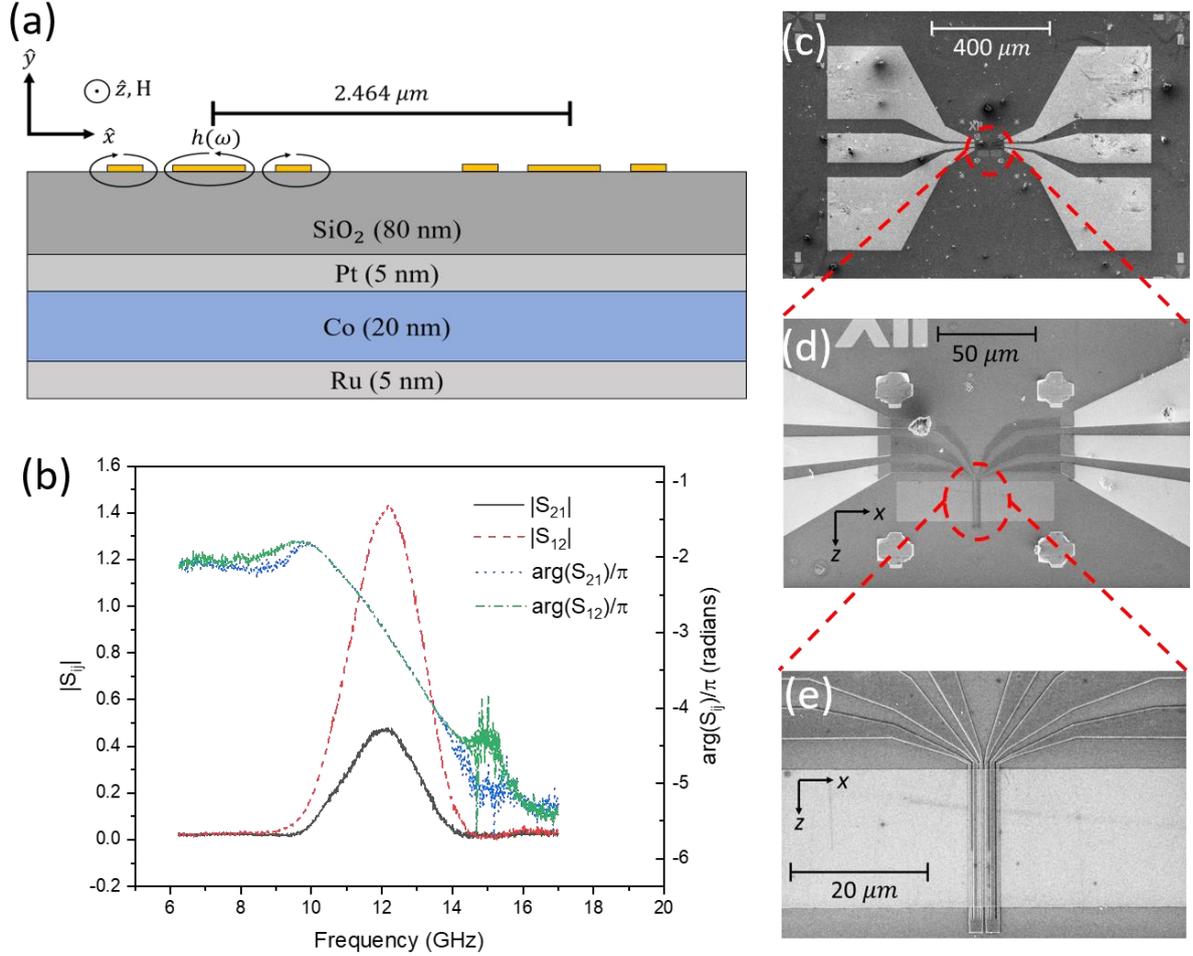

*Figure 1: (a) Tri-layer sample stack geometry. (b) Example of the amplitude and phase of the transmission characteristic ($S_{21}$, $S_{12}$) measured with the vector network analyzer. (c) Gold contact pads and feeding lines. (d) Feeding lines and spin-wave antennas on ferromagnetic strip. (e) SEM image of the spin-wave antennas.*

## 3. Experimental Setup

### 3.1. FMR

Initial characterization of the sample was conducted with ferromagnetic resonance (FMR) spectroscopy. These measurements were conducted on parts of the original continuous film which had been cut off from the main sample prior to the nano-fabrication. For the FMR, the sample was placed onto a microstrip line which was installed between the poles of an electromagnet. A microwave generator, connected to the input port of the microstrip line, was used to drive the magnetization precession in the Co layer of the tri-layer stack. The output port of the stripline was connected to a microwave diode, in order to rectify the microwave signal, and fed to a lock-in amplifier. A signal generator, producing a 220 Hz sine wave, was referenced to the lock-in amplifier and was used to drive a small modulation coil, which produced an AC magnetic field of ~10 Oe parallel with the external field of the large electromagnet. The driving microwave frequency was set constant and the externally applied magnetic field was swept over the FMR resonance, a measurement mode known as field-resolved FMR.

The in-plane (IP) FMR geometry was employed in order to supplement the Brillouin light scattering (BLS) measurements discussed in the next section. In the case of FMR, we are sensitive to the fundamental spin-wave mode with a wavenumber of 0. The BLS measurements on the other hand are sensitive to higher wavenumbers and thus together with the IP FMR one can create a better understanding of the spin-wave dispersion relation. This is discussed in the BLS results section. Additionally, in-plane angle resolved FMR was used to extract any in-plane anisotropy which may be present in the continuous film.

We also used both out-of-plane (OOP) and IP FMR as a means of extracting the magnetic losses in the system. From the FMR half-width at half maximum (*HWHM*), which is a magnetic field linewidth, we were able to obtain an estimate of the magnetic losses in the system by plotting the *HWHM* versus the microwave driving frequency (*f*) and fitting the data with a linear function,

$$HWHM(f) = s \cdot f + \frac{\Delta H_0}{2}, \tag{1}$$

where $s = \frac{2\pi \alpha_G}{\gamma}$ is the slope, $\alpha_G$ is the Gilbert damping constant, $\gamma$ is the gyromagnetic ratio, and $\Delta H_0/2$ is the inhomogeneous linewidth broadening. Note, the theoretical model includes the magnetic losses as an additional imaginary component to the magnetic field and thus we only need to include the *HWHM*. To this end, we do not need to extract $\alpha_G$ and can simply use the fitting function parameters of Eq.(*1*) to model the magnetic losses in the system.

The OOP FMR configuration was also used as a first approximation of the effective magnetization ($4\pi M_{eff}$) and the gyromagnetic ratio ($\gamma$). These may be extracted by performing field resolved FMR measurements at various driving frequencies, $f$, and fitting the resultant $f$ vs $H_{res}$ curve with the well known out-of-plane Kittel equation [21],

$$f = \frac{\gamma}{2\pi}(H_{res} - 4\pi M_{eff}). \tag{2}$$

### 3.2. BLS

Brillouin light scattering (BLS) was employed to probe the spin-wave frequencies for spin-waves propagating in opposite directions. The BLS process can be expressed as a scattering of photons and magnons (quanta of a spin wave). There are two cases of interest, the creation of a magnon known as the Stokes process, and the annihilation of a magnon, known as the anti-Stokes process. During the BLS spectroscopy process, a laser light with wavelength $\lambda$ is focused onto the sample surface at some incident angle $\theta$ from the sample normal. The incident photons scatter inelastically from magnons in the system and the backscattered photons are detected by a photo-detector. In this backscattering geometry, the laser wavelength and incident angle determine the wavenumber of the magnons present in the inelastic scattering process,

$$k = 2\left(\frac{2\pi}{\lambda}\right)\sin(\theta). \tag{3}$$

In order to determine the wavelength and ultimately the frequency of the magnons (spin waves), a tandem Fabry-Pérot interferometer (TFPI) is employed. Through use of the interferometer, the frequency difference between the incident photons and backscattered photons can be measured, and when the backscattered photons result from a scattering with a magnon, this frequency difference is equivalent to the frequency of the respective magnon. In the case of the Stokes process, the creation of the magnon requires energy which is supplied by the incident photon resulting in the backscattered photon having a lower frequency than the incident photon, and thus the Stokes peak sits on the left hand side of the BLS spectrum. The reverse is true for the anti-Stokes process where the annihilation of the magnon results in a transfer of energy from the magnon to the photon, and hence the backscattered photon has a larger frequency than the incident photon. Because of this, it is possible to measure the frequency of both the forward (+$k$) and backward (-$k$) propagating spin waves in the system with a single BLS measurement.

In this work a laser with a wavelength of 532 nm was used at three different incident angles, $\theta = 20°, 40°, 60°$, corresponding to wavenumbers of, $k = 8.08\ \mu m^{-1}, 15.18\ \mu m^{-1}, 20.46\ \mu m^{-1}$, respectively. By measuring the frequency of spin waves for different wavenumbers, it is possible to recreate the dispersion relation of the spin waves.

### 3.3. PSWS

In order to excite and detect spin waves in the ferromagnetic strips via the CPL spin-wave antennas a 2-port vector network analyzer (VNA) was employed. Each port of the VNA was connected, via 50 Ohm microwave transmission cables, to a three-pronged microwave probe (picoprobe), which was attached to a translation stage. The central pin of the transmission cables was connected to the central probe of the picoprobe and the shield of the transmission cable was connected to the two out 'ground' probes of the picoprobe. The picoprobes could then be carefully lowered to make contact with the gold pads on the sample surface, which in turn are connected to the small CPL antennas on top of the ferromagnetic strip, as discussed in the fabrication section. The sample itself was sitting between the poles of an electromagnet such that the externally applied magnetic field was applied perpendicular to the ferromagnetic strip and parallel with the CPL antennas, to conform to the Damon-Eshbach geometry in which magnetostatic surface waves (MSSWs) are excited. The VNA is sensitive to both the amplitude and phase of the transmitted signal, in both directions, and the reflected signal of the entire system attached to its two ports. From those measurements, the scattering $S$-parameters of the system are automatically extracted by the instrument. Here, "the system" refers to everything connected between the two ports on the front panel of the VNA. In order to narrow the sensitivity of the VNA to the devices on the sample itself, a standard calibration, using the 85052D 3.5mm SMA calibration kit, was performed to remove the contributions of the feeding microwave transmission lines. The contribution of the feeding lines and CPL antennas remain present in the measurements as a coplanar-line based calibration standard would be required to remove these extra contributions, which we did not have access to.

During the measurement process, the externally applied magnetic field is set to a constant value and the frequency generated by the VNA is swept. The amplitude and phase of the $S$-parameters ($S_{21}$, $S_{12}$, $S_{11}$, and $S_{22}$) are measured by the VNA as a function of frequency. In order to ascertain the contribution of the spin wave to the $S$-parameters, a background measurement must be taken at a magnetic field strength where, for the chosen range of the frequency sweep, there is no excitation of spin waves. This background signal must then be removed from the signal obtained when probing the spin waves by a process known as de-embedding. The de-embedding process is explained in the Appendices. Two processes of de-embedding were explored in order to isolate the spin-wave signal from the contributions of the feeding lines. The first process isolates the spin-wave signal from the contributions of the feeding lines themselves, however, there remains an undetermined scaling factor. The factor was found to be a rather weak function of frequency; and thus, the shape of the frequency dependencies of the $S$-parameters is preserved. This process is simple and requires less auxiliary measurements. The second process requires an extra reference sample for completion of the process, however, it does not leave any undetermined scaling factor.

For the sample, the $S$-parameters were measured as a function of frequency for several magnetic field values.

## 4. Results

### 4.1. FMR

Due to the large parameter space of the sample devices, we initially characterized the sample with FMR in order to gain some information on the magnetic properties of the sample. In Figure 2, we show the FMR results. From Figure 2(a), we extracted the effective saturation magnetization, $4\pi M_{eff}$ = 15043 G and the gyromagnetic ratio, $\frac{\gamma}{2\pi}$ = 2.87 MHz/Oe. The extracted effective magnetization and gyromagnetic ratio from IP FMR results, seen in Figure 2(b), are significantly different from those extracted from OOP FMR and we believe this to be an artifact of the IP geometry. Since the IP Kittel equation is non-linear it is more difficult to fit, and requires more measurements to be taken at higher frequencies in order to obtain accurate results [22]. We also investigated IP FMR in order to check the magnetic losses of the sample in the geometry for which the PSWS were performed. We find that the magnetic losses are higher in the IP FMR configuration by ~1.5 times, and since this is the same configuration as for MSSWs we chose to use the magnetic losses extracted from the IP FMR measurements. From the IP FMR *HWHM* we extracted, $\alpha_G = 0.016$ and $\frac{\Delta H_0}{2} = 3.98$ Oe. In addition, we performed in-plane angle resolved FMR measurements from which we found that the sample contained a bulk in-plane easy axis anisotropy with an effective field strength of $H_{ub} \sim 40$ Oe. The FMR measurements were complimented with magnetometry measurements from which the saturation magnetization was found to be, $4\pi M_s = 17000$ G.

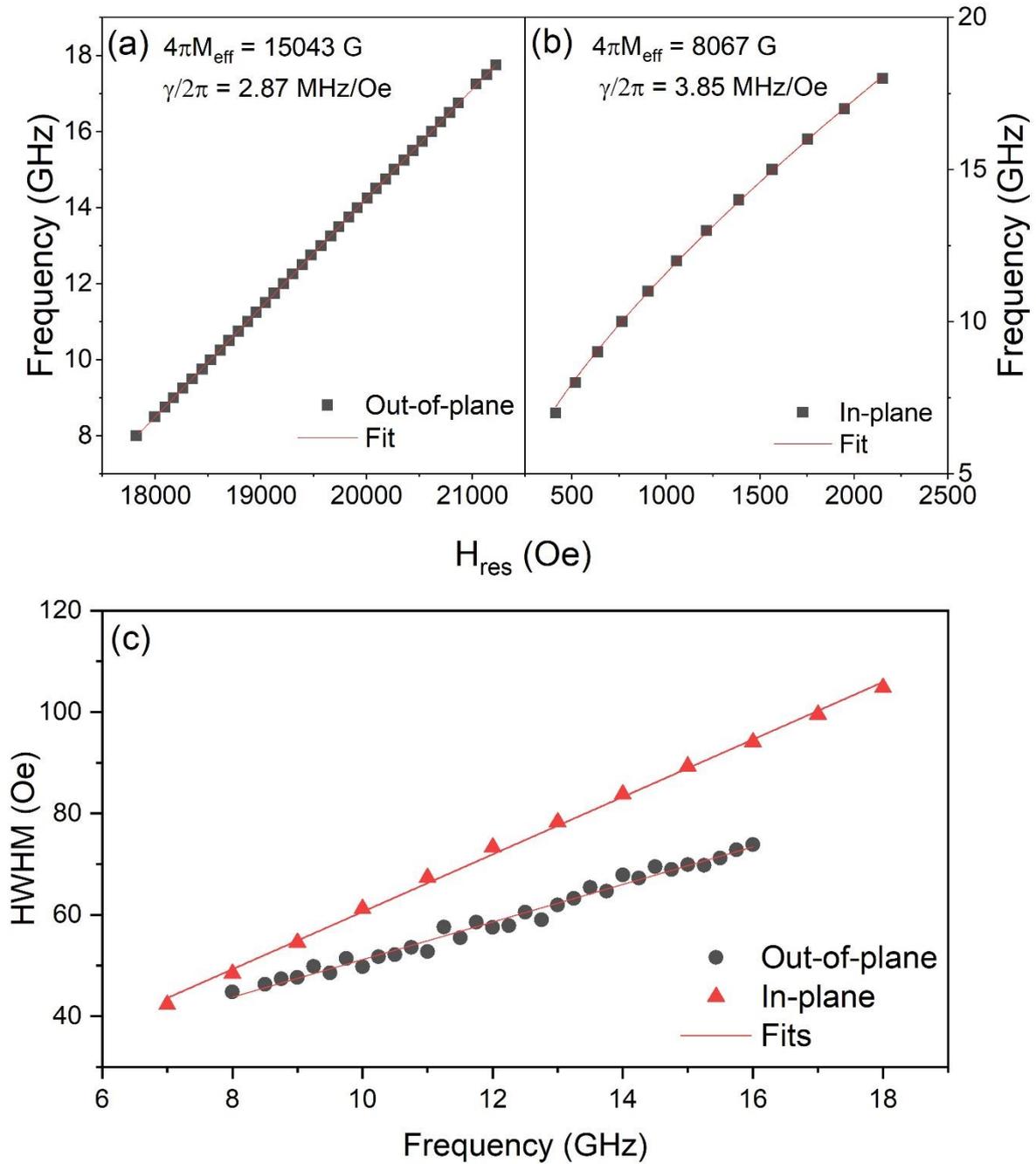

*Figure 2: (a) Microwave driving frequency vs. out-of-plane (OOP) FMR resonance field. (b) Microwave driving frequency vs. in-plane (IP) FMR resonance field. (c) OOP (black circles) and IP (red triangles) half-width at half-maximum (HWHM) of the FMR peak vs. microwave driving frequency. Red line are linear fits to the experimental data.*

### 4.2. BLS

BLS measurements were performed on the samples in order to experimentally determine the dispersion relation of the samples. Figure 3 shows the BLS stokes and anti-stokes peaks for the //Ru(5)/Co(20)/Pt(5) sample at an angle of 20 degrees from the sample normal. At this angle, one is sensitive to spin-waves with a wavenumber of 8.1 rad/$\mu$m. Two additional angles of 40 and 60 degrees, sensitive to spin-waves with wavenumbers of 15.2 and 20.5 rad/$\mu$m respectively were also considered and these BLS spectra were obtained for an externally applied, in-plane magnetic field of 3000 Oe. Another BLS spectrum was obtained at 500 Oe and an angle of 20 degrees. To determine the peak positions of the BLS peaks, the BLS spectra were first smoothed and the local maximum of the smoothed data was determined. Based on the known resolution of the employed BLS setup, an uncertainty of ~500 MHz was associated with the position of the spin wave peaks. This uncertainty stems in part from the spectral recording (i.e. the spectrum is broken up into 550 equal bins) and from the finesse of the interferometer. The frequency non-reciprocity of the Stokes and anti-Stokes peaks was found to be non-zero, but smaller than the combined uncertainty of two peaks, for all angles of investigation. As a result, it will be assumed that the DMI interaction is zero, and that any surface PMA contributions at the two interfaces, Ru/Co and Co/Pt, are identical, that is we have a fully magnetically symmetric trilayer stack.

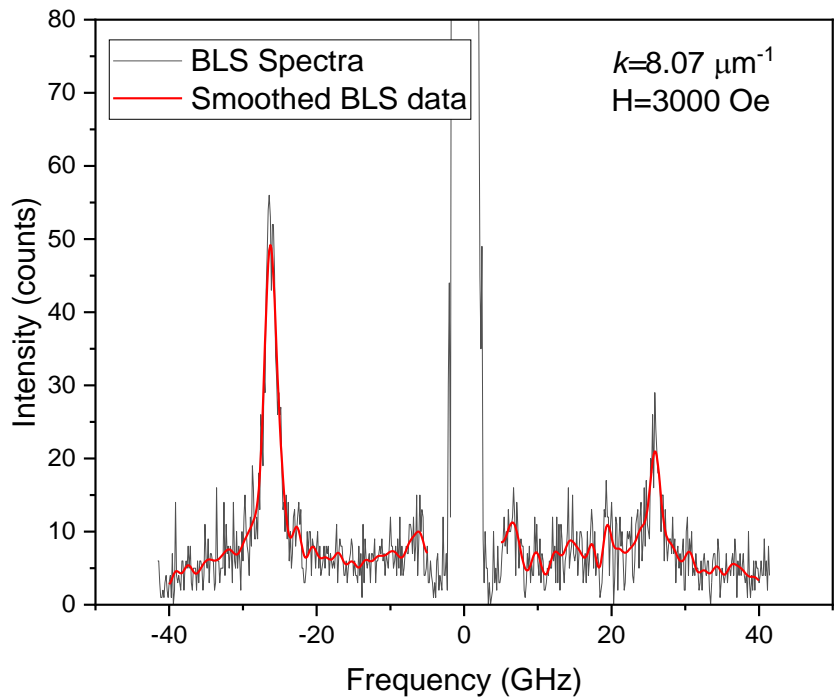

*Figure 3: Black-solid line: BLS spectra obtained for an incident laser light at 20 degrees from the sample normal and an externally, in-plane applied magnetic field of 3000 Oe. Red-solid line: 5 point fast Fourier transform smooth of the raw BLS data.*

From the Stokes peaks of BLS spectra obtained at 500 Oe we extracted the frequency of the fundamental mode (FM) as well as the frequency of the 1st standing spin-wave mode (1st SSWM). These are shown by the data points in Figure 4(a). In a similar manner, the frequencies of the FM at 3000 Oe could be determined and is plotted in Figure 4(b). The experimentally determined frequencies of the FM and 1st SSWM were fitted with a secondary numerical model (note this numerical model only models the dispersion relation and is not the main model developed in [20]) in order to determine the saturation magnetization $4\pi M_s$, the gyromagnetic ratio $\gamma$, the exchange constant $A$, and the surface anisotropy constants $K_{u1}$ and $K_{u2}$ at the Ru/Co and Co/Pt interfaces respectively. The $k=0$ point in Figure 4(b) was obtained from the in-plane FMR measurements. In order to obtain the resonance frequency of FMR mode at an externally applied field of 3000 Oe we exploited the fact that on a small frequency range the $f$ vs $H_{res}$ relationship is almost linear (here $H_{res}$ is the magnetic field at resonance). Thus, by measuring field resolved FMR for 5 frequencies around the desired frequency, $f(H_{res}=3000)$, and fitting the resultant $f$ vs $H_{res}$ data with a straight line we were able to interpolate the frequency of the fundamental mode at 3000 Oe, $f(H=3000)$.

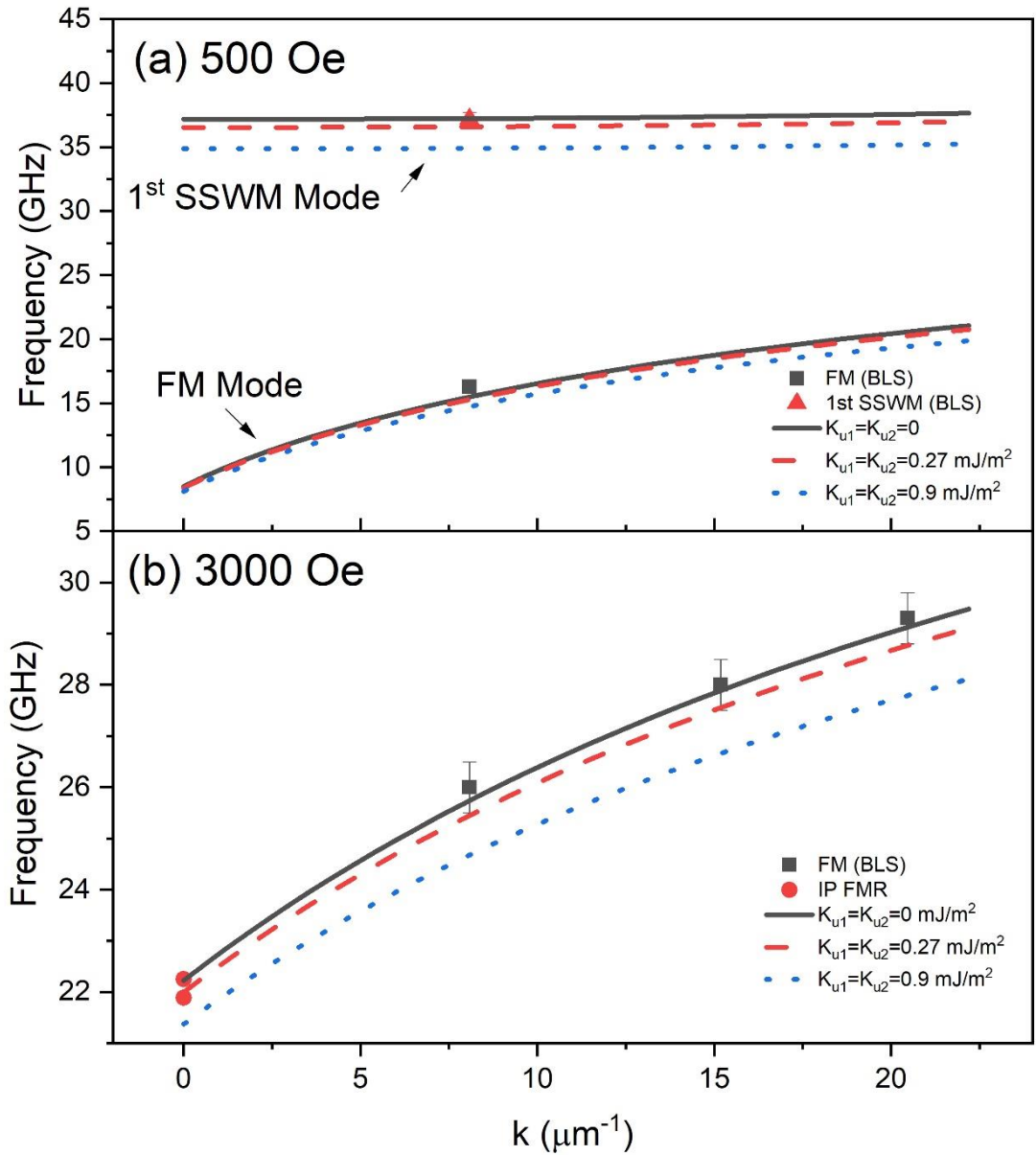

*Figure 4: (a) Fundamental mode (FM) and 1st standing-spin-wave mode (1st SSWM) obtained from BLS spectra at an externally applied field of 500 Oe. Black-square: BLS FM frequency; red-triangle: BLS 1st SSWM frequency. (b) FM at 3000 Oe. Black squares: BLS FM frequency; red circle: k=0 determined from in-plane FMR measurements. For both panels we have, black solid-lines: numerical fits for $K_{u1} = K_{u2} = 0$ $mJ/m^2$; red dashed-lines: numerical fits for $K_{u1} = K_{u2} = 0.27$ $mJ/m^2$; blue dotted lined: numerical fits for $K_{u1} = K_{u2} = 0.9$ $mJ/m^2$.*

Using $4\pi M_s$=17000 G, as deduced from SQUID measurements, the material parameters from FMR we find the parameters which result in the fits in Figure 4 are as follows, $\frac{\gamma}{2\pi}$=2.87 MHz/Oe and $A$=1.78×10⁻

[11] J/m. We also included the 40 Oe of bulk in-plane uniaxial anisotropy in the fits to the dispersion. However, since the orientation of this anisotropy is not known with respect to the applied external field of the BLS measurements, the angle between the two contributions acted as an additional free parameter. We find that the best agreement, for the case of no surface PMA, arises when the in-plane anisotropy is applied at 45° to the external magnetic field. This case is equivalent to having no in-plane anisotropy at all. We also checked several values of surface PMA strength at the two interfaces and find that the best agreement between the BLS data and the numerical dispersion arises when there is no PMA present at either interface, i.e. $K_{u1} = K_{u2} = 0$ mJ/m². If one considers the two interfaces to have the same PMA strength and uses a value of $K_{u1} = K_{u2} = 0.9$ mJ/m², as found in the literature [23], then one obtains the dispersion relation given by the blue-dotted line in Figure 4. This dispersion is significantly downshifted from the BLS data, suggesting this PMA is much stronger than what is present in the sample. The largest value of surface anisotropy that can be associated to each interface, whilst remaining within the uncertainties of the BLS data points, is $K_{u1} = K_{u2} = 0.27$ mJ/m², shown by the red-dashed lines in Figure 4. From the OOP FMR measurements however, one would expect a significant surface PMA contribution resulting in the large difference between $4\pi M_{eff} = 15043$ G and $4\pi M_s = 17000$ G. This effective PMA field would be around $H_u \sim 2000$ Oe resulting in $K_{u1} = K_{u2} = 1.35$ mJ/m² which would result in an even lower frequency shifted dispersion than that of the blue dotted curve in Figure 4. It may be the case that there is a fourth order magnetic anisotropy present in the sample. This anisotropy may affect the effective magnetization differently for the in-plane and out-of-plane magnetized sample. For Co/Pt interfaces a small fourth order magnetic anisotropy contribution has been suggested in the literature [24] and we may speculate that the Co/Ru interface also has a fourth order anisotropy. The Co/Ru interface is not well understood and thus it may well be the case that a fourth order magnetic anisotropy present at this interface is sufficiently large such that it may explain the discrepancy of the effective magnetization extracted from OOP FMR and the BLS data. As a result, we opted to disregard the value of $4\pi M_{eff}$ extracted from OOP FMR and utilized instead just the $4\pi M_s$ extracted from magnetometry and the surface PMA values which resulted in the best fits of the BLS data, namely $K_{u1} = K_{u2} = 0$ mJ/m². These parameters were used in the developed theoretical model in order to fit the transmission $S$-parameters of the experimental data.

### 4.3. $S_{21}$ and $S_{12}$

The optimal magnetic, electrical, and geometric parameters of the investigated device are displayed in Table 1.

*Table 1: Optimal magnetic, electric, and geometric parameters of the device investigated.*

| | | | | | |
|---|---|---|---|---|---|
| $w$ (nm) | 810 | $d_1$ (nm) | 5 | $\varepsilon_s$ | 3.8 |
| $w_g$ (nm) | 450 | $d_2$ (nm) | 5 | $\rho_{Al}$ ($10^{-8}$ Ω m) | 2.65 |
| $\Delta_g$ (nm) | 180 | $d_s$ (nm) | 130 | $K_{u1}$ (mJ/m²) | 0 |
| $l_s$ (μm) | 20 | $4\pi M_s$ (G) | 17000 | $K_{u2}$ (mJ/m²) | 0 |
| $l_d$ (μm) | 2.464 | $\gamma$ (MHz/Oe) | 2.87 | $D_1$ (pJ/m) | 0 |
| $\alpha_G$ | 0.016 | $H$ (Oe) | 612 | $D_2$ (pJ/m) | 0 |
| $\frac{\Delta H_0}{2}$ (Oe) | 7.97 | $\sigma$ ($10^7$ S/m) | 1.7 | $H_{ub}$ (Oe) | 40 |
| $A$ ($10^{-11}$ J/m) | 1.78 | $\sigma_1$ ($10^7$ S/m) | 1.409 | | |

| L (nm) | 20 | $\sigma_2$ (10$^7$ S/m) | 0.952 | |

Figure 5 shows the experimentally determined $S'_{21}$ and $S'_{12}$ parameters overlaid with the theoretically simulated $S'_{21}$ and $S'_{12}$ parameters for the //Ru(5)/Co(20)/Pt(5) sample. For the simulation, the optimal parameters described in Table 1 were used and an in-plane applied external field of 612 Oe was assumed. The $S'$-parameters correspond to the raw $S$-parameters after undergoing the de-embedding procedure outlined in Appendices A and B.

Note that the theory from [20] assumes that the sample is continuous in its plane. In reality, the sample represents a strip of finite length in the direction $x$ and finite width in the direction $z$. This makes the internal static magnetic field inside the strip different from the applied one. It must also modify slightly the dynamic demagnetizing (dipole) field created by the traveling spin wave. If the strip length were much larger than its width, we could base our theory on the Green's function of the dipole field of a guided spin-wave on a strip waveguide [25]. Unfortunately, that Green's function is two-dimensional, and therefore, using it would increase the simulation time by orders of magnitude. Furthermore, it is incompatible with our Telegrapher Equations approach. However, it was previously shown [26] that the main effect of the strip geometry is that the internal static field is smaller than the applied one, provided that the strip width is larger than the spin wave attenuation length.

Therefore, in order to be able to compare results of our modeling with the experiment, we introduce small corrections to the theory. We add the field of a uniformly magnetized rectangular prism to the effective field entering the Landau-Lifshits Equation. We do this by employing the effective demagnetizing factors for the prism. We calculate them using Eq. (1) in Ref. [27]. Thus, for the ferromagnetic strip with the following dimensions: $x = 80$ $\mu$m, $z = 20$ $\mu$m, and $y = 20$ nm, we calculate the following demagnetizing factors: $N_x = 0.0006$, $N_z = 0.0026$, and $N_y = 0.9968$.

Note that the $N_z$ factor acts on the static magnetization, and thus decreases the static magnetic field entering the linearized Landau-Lifshits equation. Adding it to the theory is a reasonably accurate approach. Conversely, the $1-N_y$ factor and the $N_x$ factor act on the dynamic magnetization. Therefore, this approach is acceptable for vanishing spin-wave wavenumbers, but loses its accuracy with an increase in the spin-wave wavenumber. Furthermore, the dipole field generated by a spin wave, with a wavenumber $k$, scales as $kL/2$, where $L$ is the thickness of the ferromagnetic layer. This yields an "effective demagnetizing factor" for the spin wave for the center of the $S_{21}$ transmission band ($k=3.1\times10^6$ m$^{-1}$) of 0.031. This is significantly larger than $N_x$ and $1-N_y$. Thus, the effect of the in-plane static demagnetizing factors is negligible for large spin-wave wavenumbers, but provides correction for small wavenumbers.

The ultimate effect of the inclusion of the $N_x$ and $1-N_y$ factors is a slight correction of the curvature of the frequency vs. applied field dependence of $S_{21}$ on the band (which will be shown in Figure 6 later on). Furthermore, if the uniaxial magnetocrystalline anisotropy is described with effective demagnetizing factors of anisotropy, it is indistinguishable from the shape anisotropy [28]. Therefore, a good fit of the curvature can be interpreted either as the effect of shape anisotropy or the presence of extra uniaxial anisotropy of the same strength.

Again, we also needed to include the 40 Oe of bulk in-plane uniaxial anisotropy into the theoretical model and opted to apply this effective field in the $x -$ direction, namely along the ferromagnetic strip. This is a possible scenario, meaning that the patterning of the strip results in the magnetocrystalline uniaxial anisotropy axis wanting to align itself with the easy axis of the strip shape anisotropy.

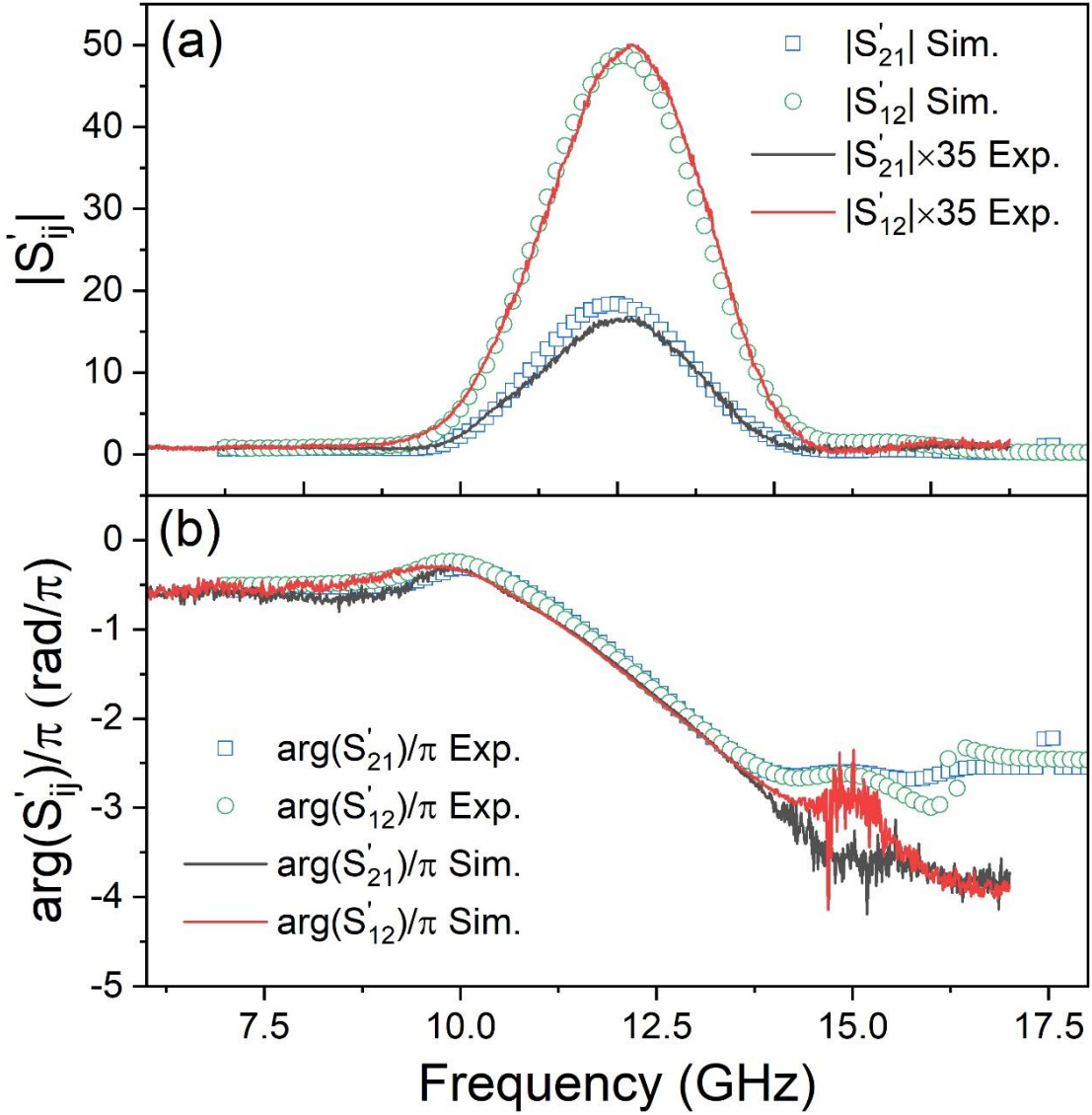

*Figure 5: (a) Amplitude and (b) Phase of the de-embedded transmission, $S'_{ij}$,-parameters of the experiment and theoretical simulation at an externally in-plane applied magnetic field of 612 Oe. Black solid-line: Experimental $S'_{21}$. Red dashed-line: Experimental $S'_{12}$. Blue dotted-line: Simulated $S'_{21}$. Green dash-dotted-line: Simulated $S'_{12}$. Note, the experimental amplitude curves in panel (a) have been scaled by a factor of 35 such that they are at the same magnitude as the simulated curves.*

Both the theoretical and experimental transmission characteristic have been de-embedded following the process described in the Appendix A. After de-embedding, both the experimental and simulation curves, we find that the experimental amplitude curves are ~ 35 times smaller than the simulated amplitude curves. Thus, the experimental curves were scaled up by a factor of 35 times in order to be of similar magnitude as the theoretically simulated curves, and are shown in Figure 5(a). We expect this to be a result of parasitic direct coupling of the input microwave path to the output one, which is not accounted for in the theoretical model and is explained later. From Figure 5(a) we see that the frequency position of both the forward ($S_{21}$) and backward ($S_{12}$) curves matches very well. The theoretical $S_{21}$ and $S_{12}$ peaks in Figure 5(a) have a maximum at 11.925 GHz and 12.045 GHz respectively, which correspond to a wavenumber of ~2.8 rad/$\mu$m and ~3.0 rad/$\mu$m respectively. The experimental $S_{21}$ and $S_{12}$ peaks in Figure 5(a) have a maximum at 12.060 GHz and 12.171 GHz respectively. This suggests that the material parameters chosen for the theoretical model are very close to the true parameters of the material as is expected from our analysis of the BLS data and dispersion relation. We find that both the simulated peaks and the experimental peaks have the same trend of the frequency non-reciprocity, namely that the backward ($S_{12}$) traveling wave peak maximum has a slightly higher frequency than the forward ($S_{21}$) traveling wave one. Although the experimental and theoretical frequency non-reciprocities do not coincide perfectly (120 MHz for the simulation and 111 MHz in the experiment), they are very close. In our calculation, the Dzyaloshinskii constant for both interfaces is zero, and both surface PMA constants at the two interfaces, Ru/Co and Co/Pt, are the same (and vanishing), which should result in a perfectly symmetric dispersion relation. Therefore, the frequency difference between the transmission maxima of backward and forward waves might be induced by the process of the spin-wave excitation and reception by the CPL, as mentioned in [20].

We have set the iDMI values to zero at both interfaces as we find that the presence of realistic values of the DMI constant (<10 pJ/m) at either of the interfaces results in changes of the frequency position of the maximum of transmission which are smaller than 200 MHz and thus cannot be accurately decoupled from the uncertainty. The same is true for the surface PMA values at the two interfaces. In our simulations, we assumed an equal distribution of PMA across the two interfaces as suggested by the BLS measurements. However, the literature suggests that the surface PMA value at the Co/Pt interface may be larger than the surface PMA value at the Ru/Co interface [29].

We now turn to the phase profiles of the transmission parameters shown in Figure 5(b). The simulated phase profile and experimental phase profile have similar slopes which suggests that the theoretical model accurately accounts for the group velocity of the spin-waves in the material. If we ignore any contribution to the transmission phase due to the excitation and reception of spin waves by the antennas, then we may relate the slope of the phase profile in Figure 5(b) directly to the slope of the dispersion relation, also known as the group velocity as seen by Eq. (S18) of the online supplementary materials in [20].

Note, we do not investigate the reflection coefficients in our analysis. This is a result of the de-embedding process, described in Appendix A, after which one completely removes the reflection coefficients on a first-order approximation.

Next, we compare the results of the theoretical model with the experiment for several different externally applied magnetic field strengths. For each magnetic field value, we fit the amplitude of the $S_{21}$ and $S_{12}$ transmission peaks with an Edgeworth-Cramer peak function and then find the local maximum of this fitted function. This local maximum is taken as the nominal frequency of the excited spin waves at the corresponding applied magnetic field. This is done for both the experimental peaks and the theoretical peaks. The frequency versus applied magnetic field data is plotted in Figure 6. Note, that the simulations are calculated at slightly different magnetic fields than the experiment. However, the underlying trend is not affected and the comparisons that follow are still valid.

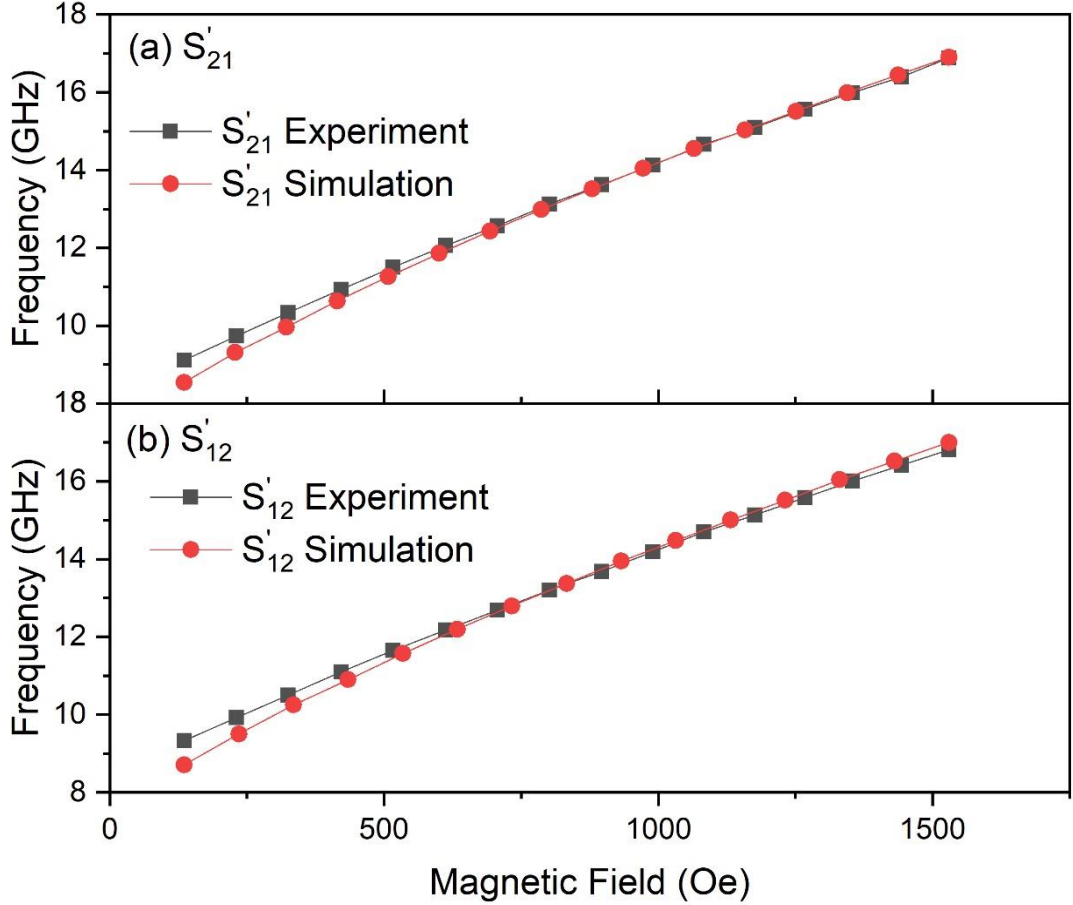

*Figure 6: The frequency of the (a) |S$_{21}$| and (b) |S$_{12}$| peak maximum positions versus the externally in-plane applied magnetic field. Black squares: theory. Red circles: experiment. Solid lines are linear curves connecting individual data points and are included solely as a guide for the eye.*

Figure 6 shows a good agreement between the theoretical simulation and the experimental measurements of the transmission peak positions at varying magnetic fields. Thus, although we originally only used the experimental data at 612 Oe to determine the sample parameters which best fit the experimental |$S_{ij}$| peaks with our theoretical simulation, we find that the simulation is in good agreement at a large range of magnetic fields. There is a small difference between the theory and the experiment, which is most pronounced for smaller magnetic fields and also stronger in the $S_{21}$ than the $S_{12}$ transmission characteristic. We believe this discrepancy arises from our simplified description of the effect of the shape anisotropy of the magnetic strip, and it is expected that we will not be able to fully recreate the experimental curvature with the numerical model for this reason. Additionally it is expected that there may be an uncertainty of ~200 MHz associated with the choice of CPL antenna geometry and antenna elevation above the trilayer stack in the numerical simulation.

As mentioned previously, there is a significant difference in the magnitudes of the amplitudes of the transmission parameters between the experiment and the simulation. One explanation for this is an extra direct parasitic coupling between the receiving and transmitting antennas which is present in the experiment but not accounted for in the theoretical model. From the final result of the de-embedding process (Eq. (A26) and Eq. (B1)), we see that the de-embedded signal is not only sensitive to the spin wave transmission but also the background direct coupling transmission. Thus, if the direct coupling transmission increases, then we may expect a decrease in the total de-embedded signal since the direct inductive coupling contribution enters in the denominator of the factors in front of the matrices on the right-hand sides of Eq. (A26) and Eq. (B1). This extra coupling may come from sections of the antennas that are not accounted for in the model. In the model, the length of the antenna is assumed to be equal precisely to the width of the magnetic strip, and the short ends are assumed not to have any spatial extension. In the real sample device, the antenna goes slightly beyond the ferromagnetic strip and the short-end pads are noticeably extended. Additionally, there may be contributions from the curved feeding lines connecting the antennas to the gold contact pads. The feeding lines are long, therefore we expect that the main parasitic coupling actually originates from them.

In order to understand the amplitude of the spin-wave transmission signal, we developed another de-embedding procedure in which we characterize the transmission and reflection parameters of the feeding lines. To do this, we measured the reflection parameter ($S_{11}$) from one of the devices for two cases. 1) Where the CPL antenna was intact and considered as a load impedance connected at the end of the feeding lines as shown in Figure 7(a). 2) Where the CPL antenna strips were each cut at the end of the feeding lines as shown in Figure 7(b). The cut was performed with a focused ion beam (FIB). For this second case, the transmission line path ends in an open load ($Z_L = \infty$). Note, that the cut was performed on a different antenna than the one used to perform the measurements in this paper. However, the antenna geometries were identical and thus we expect the resultant $S_{11}$ of the cut antenna to be a good representative of the antenna used in this study. By measuring the $S_{11}$ for these two cases and making an assumption of the characteristic impedance of the output port of the feeding line, it was possible to approximate the **ABCD** matrix of the feeding line. Note, that here the "feeding lines" refer to the cascaded network of the picoprobe as well as the Au contact pads and the Al feeding lines which connect the Au contact pads to the CPL antenna. The details of this characterization are explained in Appendix C.

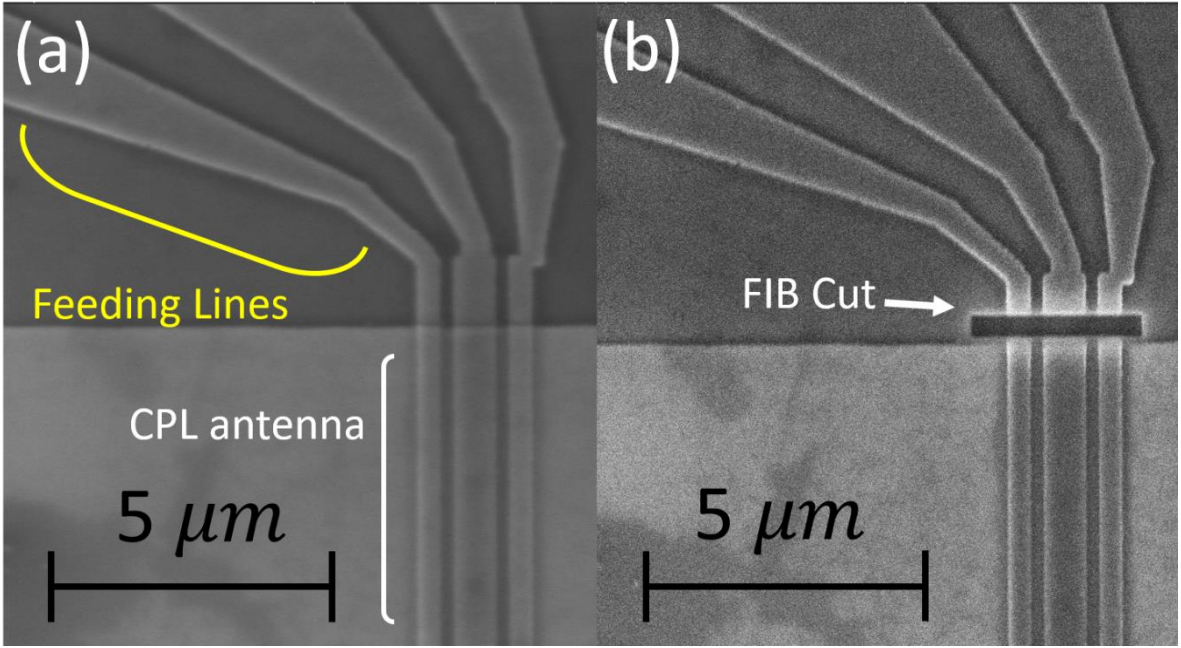

*Figure 7: (a) Feeding line loaded with the intact CPL antenna. (b) Feeding line cut at the start of the CPL antenna.*

Once the **ABCD** matrix of the feeding lines is known, then, mathematically, it is straightforward to de-embed the contributions of the feeding lines from the total measured $S$-parameters of the entire device structure (input feeding lines, spin-wave delay line, and output feeding lines). The mathematical details of this de-embedding are described in Appendix D. In Figure 8 the results of the de-embedded experimental $S^*$-parameters are compared with the $S^*$-parameters of the numerical simulation. Note, both sets of $S^*$-parameters have been background corrected as discussed in Appendix D. These corrected parameters are denoted by $S^*$.

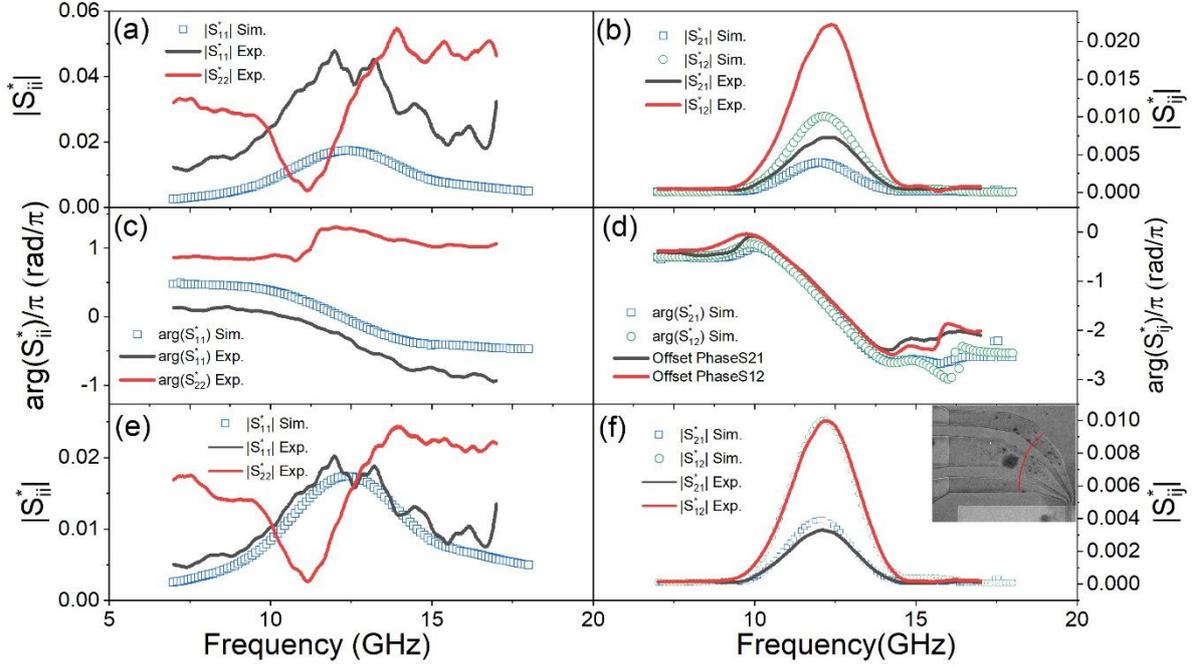

*Figure 8: Comparison between the de-embedded experimental $S^*$-parameters and the theoretically simulated $S^*$-parameters. (a) Amplitude of the reflection $S^*$-parameters. (b) Amplitude of the transmission $S^*$-parameters. (c) Phase of the reflection $S^*$-parameters. (d) Phase of the transmission $S^*$-parameters (e) Amplitude of the reflection $S^*$-parameters when extracting the $Z_c^{antenna\ port}$ (f) Amplitude of the transmission $S^*$-parameters when extracting the $Z_c^{antenna\ port}$. Inset: SEM image of the curved feeding line*

Note that in Figure 8 the simulated *S*-parameters are also background corrected by simulating the off-band background at a large field of 10000 Oe and subtracting the resultant *S*-parameters from the on-band *S*-parameters. The process of de-embedding results in a very good qualitative and quantitative agreement between the transmission $S^*$-parameters of the theory and experiment, seen in Figure 8(b). This is evidenced by the very good agreement in shape and position. The amplitude of the de-embedded experimental peak and theoretically simulated one are also very similar and differ only by a factor of ~2.2, which is a significant improvement to the factor of 35 seen for the initial de-embedding procedure. The amplitude of the reflection parameters are also in good agreement, as seen in Figure 8(a), with a similar difference in magnitudes of ~2. As shown in Appendix C, our de-embedding method relies on an assumption for a characteristic impedance of the feeding line port facing the antenna $z_c^{antenna\ port}$. In this round of de-embedding, we assumed that the port characteristic impedance is the same as the characteristic impedance of the CPL of the antenna, given that the FIB cut was across the very beginning of the antenna. However, the length of the CPL section with the same geometry on the feeding-line side of the cut (i.e. above the cut in Figure 7(b)), is quite small (about 0.5 μm), after which the CPL starts to gradually become wider. Both, the 0.5 μm size and the total length of the narrowest parts of the feeding lines, where the CPL geometry gradually changes (grey sections of CPL in Figure 1(d) and inset to Figure 8(f)), are much smaller than the wavelength of microwaves for the transmission band of our spin-wave devices (about 1 cm, if the Si substrate is taken into account). Therefore, the assumption of a well-defined characteristic impedance for the port is perhaps not very well justified.

Based on this idea, we improved the de-embedding process with one additional step. Given that the shapes of the frequency dependencies of both amplitude ($|S_{21}|$)) and phase of $S_{21}$ are in good agreement between theory and experiment, we now assumed that our theory also delivers the correct $|S_{21}|$, and that the 2.2 times difference between the theoretical and experimental de-imbedded amplitudes is due to an incorrect assumption of the value for $z_c^{antenna\ port}$. If so, by fitting the de-embedded experimental data to the theory, we extracted a value of $z_c^{antenna\ port}$ that results in the best overlap of the theoretical and experimental $|S_{21}|$. The result of the fit is shown in Figure 8(f). The value of the real and imaginary components of the characteristic impedance that we extracted from the fits ($z_c^{antenna\ port}$) is shown by the blue dotted and green dash-dotted curves in Figure 14 from Appendix C, respectively. This corresponds to ~3.75 Ohm/m of ohmic resistance per unit length $R$ and ~7.48 Ohm/m of capacitive inductance per unit length $Y$ (Eq. (11) from Appendix C). The latter was calculated for a frequency of 12.1 GHz, which corresponds approximately to the maxima of the amplitude of the transmission $S$-parameters. $Y$ is the same as we originally assumed for the port, but $R$ is smaller by 6.6 times. Note that $|z_c^{anenna\ port}|$ scales as $\sqrt{R}$, therefore, the correction in $z_c^{antenna\ port}$ is just 2.6 times.

In order to understand how these values relate to the geometry of the feeding line, we measured the widths of the ground and signal lines and the gaps between them at half the length of the bent part of the feeding line. (The inset to Figure 8(f) shows the cross-section, for which we took these measurements.) We found that the signal line is wider by 8.5 times than at the cut. The ground lines are wider by 6 and 12 times. The gaps increased by 16 and 22 times. Given that, we may expect a 6 to 10 times decrease in $R$ for this cross-section. We also calculated $Y$ using our theory of capacitance assuming a 9-times larger width of the signal and ground lines (the mean value of 6 and 12) and the 19-time increase in the gap width with respect to the sizes at the cut. We found that $Y$ for the maximum of device transmission band is 14 Ohm/m compared to 10 Ohm/m for the CPL geometry at the cut. Because $z_c^{antenna\ port}$ scales as $1/\sqrt{Y}$, the change in $\sqrt{Y}$ of 20% does not significantly affect $z_c^{antenna\ port}$.

This calculation confirms that we are dealing with some effective characteristic impedance of magnitude that is similar to what one would expect for a regular (i.e. constant cross-section geometry) CPL with the same geometry of cross-section as one for the cross-section shown in the inset to Figure 8(f). Closer to the antenna, $R$ is larger and farther from it, $R$ is smaller. Thus, "on average" one would expect $R$ to be close to one for the cross-section from the inset of Figure 8(f), and, given that $\sqrt{Y}$ does not vary much along the bent part of the feeding line, we may expect the effective characteristic impedance to be similar to one for a regular CPL with the geometry of the shown cross-section. This reasoning is in agreement with the extracted $z_c^{antenna\ port}$ value.

Note that this correction also resulted in a much better agreement between the theoretical $S_{11}$ and the de-embedded experimental $S_{11}$. (Figure 8(e)). This is one more evidence that our theory delivers an accurate value of the magnitudes of the $S$-parameters and that the method of the effective characteristic impedance is valid.

The slopes of the phases of both transmission and reflection are similar as seen in Figure 8 (c) and (d). As a result, we claim that our initial de-embedding procedure, shown in Figure 5 and explained in Appendix A, acts as a good alternative when the characteristics of the feeding lines are not known exactly. Using this original de-embedding procedure outlined in Appendix A, one obtains a good agreement in the shapes and frequency positions of the de-embedded experiment and simulated signal. The only discrepancy then remains the difference in amplitudes which can be explained as a larger background direct coupling in the experiment. This difference in coupling can be extracted by simply rescaling the simulated data to overlap with the experiment.

Thus, if one is not interested in the value of the magnitude of the spin-wave channel contribution to $S_{21}$ and $S_{12}$, but only in the peaks' shapes and positions, the method of de-embedding from Appendix A is more suitable. Conversely, if extracting the amplitudes of the peaks is necessary, and if the background coupling is not smooth over the investigated frequency range, one needs to employ a more experimentally involved procedure, such as the destructive method from Appendix D. One then obtains a value that is close to the real one if a realistic assumption for the $z_c^{antenna\,port}$ is made. Assuming that it is equal to the characteristic impedance of the antenna is already a good approximation, as the error of just 2.2 times (6 dB) is not very significant if the goal is to estimate $S_{11}$ with an accuracy of $\pm 5$ dB, which may often be the case.

Let us now discuss the effects that the CPL antenna geometry and elevation have on the theoretical transmission characteristics. The elevation refers to distance between the CPL antenna and the ferromagnetic strip in the y-direction. We find that the effects of the geometry and elevation are non-negligible, and since it is difficult to determine the exact antenna geometry and elevation of the antennas, we attribute an uncertainty of ~200 MHz with the frequency of the maximum of the theoretical transmission characteristic.

As mentioned, the antenna geometry we employed in the theoretical simulation was determined by estimating the CPL strip widths from the SEM image in Figure 1(e). This "measured" geometry had thicker signal and ground strips, and thinner inter-strip gaps than the "nominal" geometry (given by the lithographic mask). In order to check the effects of this discrepancy we also checked the theoretical simulation at an applied in-plane magnetic field of 612 Oe for CPL antenna geometries matching the "nominal" geometry. The antenna geometry in this case was, $w$=648 nm, $w_g$=324 nm, $\Delta_g$=334 nm. Figure 9 shows comparisons between the two antenna geometries, where we refer to the geometry given by the lithography mask as the "nominal", and the geometry given by the SEM image as "measured".

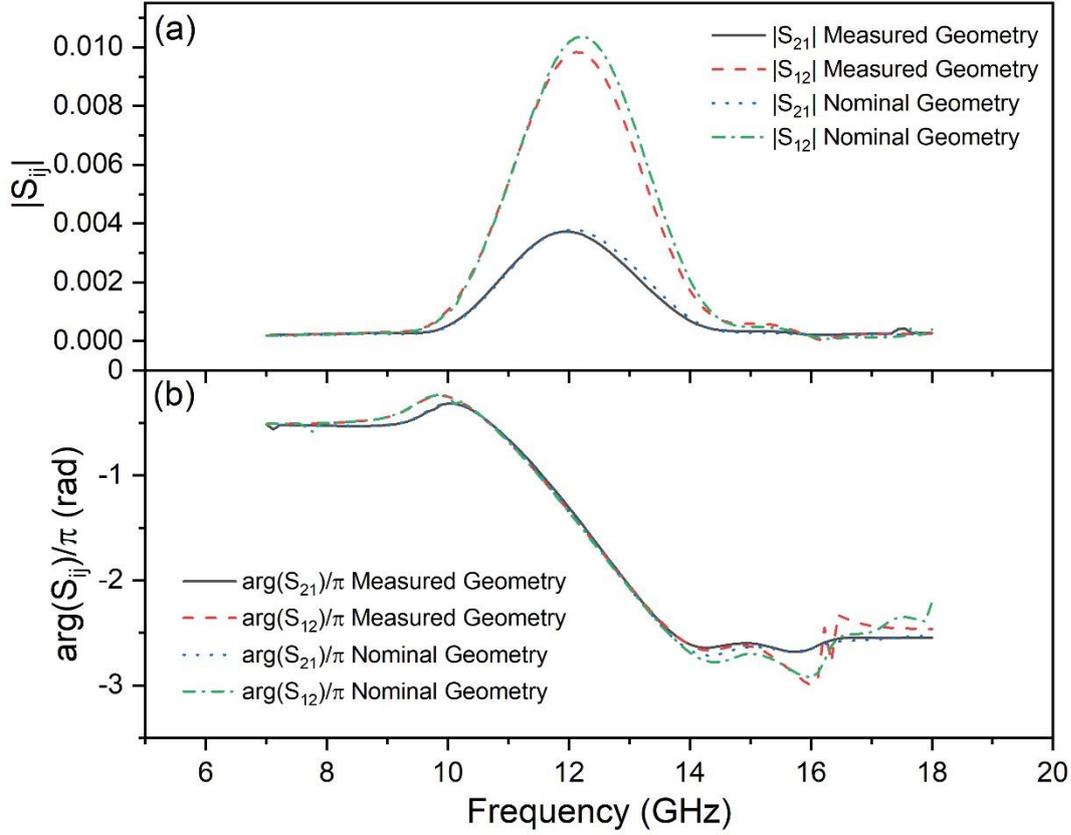

*Figure 9: (a) Amplitude of the raw transmission parameters. (b) Phase of the transmission parameter for the two antenna geometries (nominal and measured). Black-solid line: $S_{21}$, measured geometry; red dashed-line: $S_{12}$, measured geometry; blue dotted-line: $S_{21}$, nominal geometry; green dash-dotted-line: $S_{12}$, nominal geometry.*

The main effect of the antenna geometry is on the amplitude of the transmission characteristic as can be seen in Figure 9(a). There is a slight change in the frequency of the transmission peaks, which is expected since the antenna geometry determines the nominal transmission band of the antenna, that is, it defines the nominal wavenumber that the antenna may excite and be receptive to. As a result, changing the antenna geometry would result in a different nominal wavenumber and thus one would expect a different nominal spin-wave frequency. The change in frequency between the two antenna geometries is ~60 MHz for the $S_{21}$-parameter and ~90 MHz for the $S_{12}$-parameter. Additionally, there is a difference in the amplitude heights which may be a result of the more localized coupling for the thinner "nominal" antenna geometry. From the phase profiles in Figure 9(b) we notice little to no difference. This suggests that although the individual strips of the antennas differ significantly, the total spin wave propagation distance between excitation and reception stays approximately the same and thus leads to the same phase accumulation. This may be expected since the center to center distance between the excitation and receiving antenna was kept constant at 2.464 μm for both antenna geometries, and furthermore, both antenna geometries have almost identical total widths. The total widths correspond to: $w+2w_g+2\Delta_g$ and are 1.964 and 2.07 um for the "nominal" and "measured" antenna geometries respectively. Thus, changing the individual strips and inter-strip gaps of the antenna geometry on the order of 100 nm does not have a substantial impact on the shape of the transmission curve and only the frequency and amplitude are slightly affected.

Next, we check how the elevation above the trilayer stack of the CPL antennas affects the theoretical transmission characteristics. This is important as one of the main assumptions made in the theoretical model is the presence of infinitely thin CPL antennas, which in the real system is not the case. In the real system, the antennas have a finite thickness of ~100 nm. Thus, the height of the antennas in the theoretical model is somewhat arbitrary and the model assumes the antennas are sitting at a height which corresponds to the halfway point of the real finite thickness antennas. However, it may be more suitable to place the infinitely thin antennas directly on top of the dielectric spacer layer ($SiO_2$), at ~$y$=105 nm, with $y$=0 nm corresponding to the Ru/Co interface. This is justified by the fact that the antenna thickness (in the direction $y$) in our experiment is 1/6 of skin depth for aluminum for our frequency range. This implies that the microwave current is distributed almost uniformly through the antenna thickness.

The lower elevation may be modeled by simply decreasing the thickness of the dielectric spacer layer such that it accounts for the nominal dielectric spacer thickness (80 nm) only. Figure 10 shows how the simulation with the thinner spacer layer (80 nm) compares to the typical thicker spacer layer (130 nm). In addition we also checked the numerical model when the CPL antennas were placed very close to the top of the trilayer stack, i.e. we reduced the spacer thickness to just 1 nm.

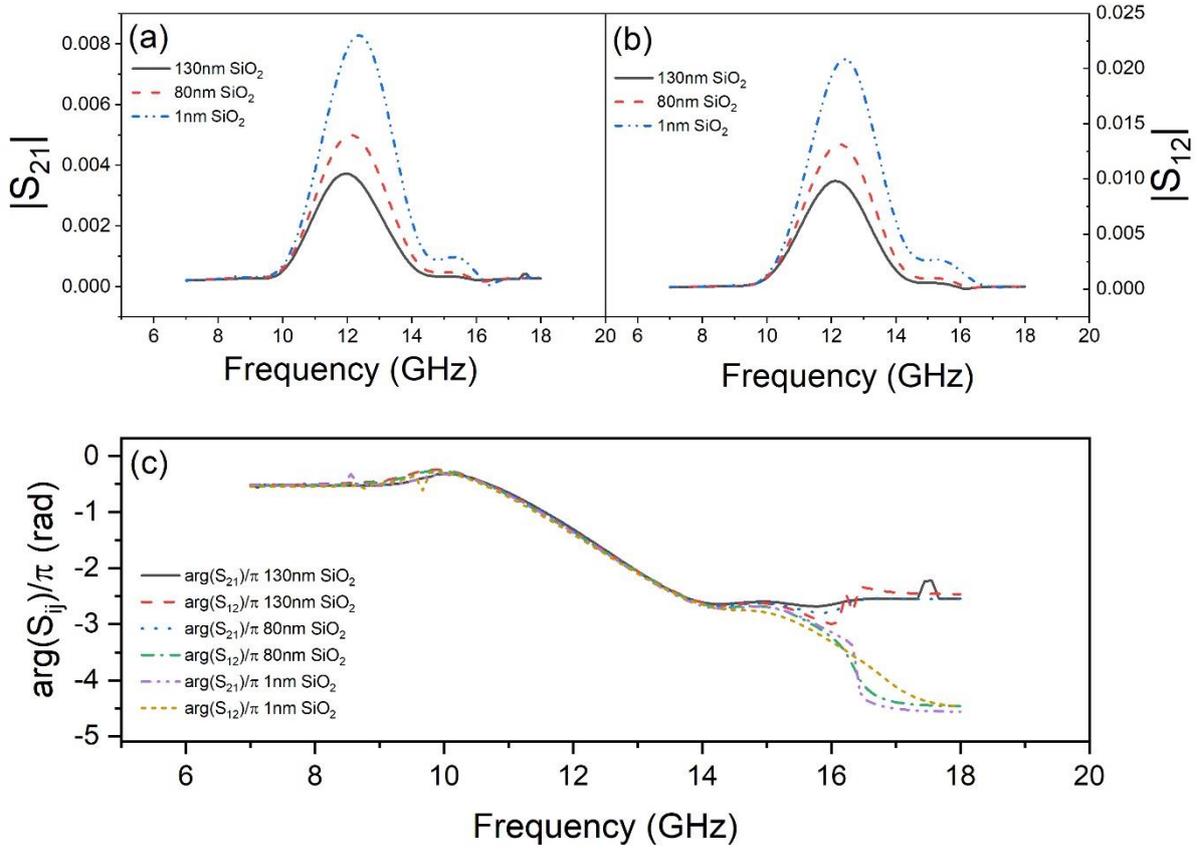

*Figure 10: (a), (b) Amplitude of the de-embedded theoretical $S_{21}$ and $S_{12}$ transmission characteristics, respectively. Black solid-line: 130nm $SiO_2$; red dashed-line: 80nm $SiO_2$; blue dash-dotted-line: 1nm $SiO_2$. (c) phase of the de-embedded theoretical transmission characteristic. Black solid-line: $S_{21}$, 130nm $SiO_2$; red dashed-line: $S_{12}$, 130nm $SiO_2$; blue dotted-line: $S_{21}$, 80nm $SiO_2$; green dash-dotted-line: $S_{12}$, 80nm $SiO_2$; purple dash-dot-dotted-line: $S_{21}$, 1nm $SiO_2$; orange short-dashed-line: $S_{12}$, 1nm $SiO_2$.*

By decreasing the $SiO_2$ spacer layer thickness, and ultimately the elevation of the CPL antennas above the trilayer stack, one increases the amplitude of the transmission characteristic of the spin waves. This is expected, since coupling of the antenna current to a spin wave with a wavenumber $k$ scales as $\exp(-ky)$, where $y$ is the antenna elevation, and thus the coupling of the CPL antenna to the ferromagnetic layer increases as the elevation of the CPL antenna is decreased (see e.g. [20] and Eq.

(55) in [30]). Similar to changing the antenna geometry we also see a change in the frequency of the maximum of the transmission when changing the $SiO_2$ spacer thickness. In this case we see that decreasing the spacer thickness results in an upwards shifting of the frequency. On average for the forward and backward propagation directions we find that the frequency shifts upwards by ~120 MHz when decreasing the spacer from 130 nm to 80 nm and shifts upwards by ~170 MHz when decreasing the spacer thickness from 80 nm to 1 nm.

Thus, we find that realistic uncertainties of both the antenna geometry and the $SiO_2$ spacer layer thickness may result in shifts of the frequency of the maximum of transmission by ~200 MHz. Given that the physical CPL antennas on the sample device have non-negligible thicknesses and that the thicknesses of the deposited layers may have a cumulative uncertainty of several nanometers associated with them, it is difficult to determine at which elevation one should set the antennas to in the theoretical model. Additionally, it is difficult to determine the true CPL antenna geometry and we expect that the "measured" geometry chosen for the simulations may have an uncertainty of several tens of nanometers when compared to the true geometry. Thus, both the uncertainty in antenna elevation as well as in antenna geometry may result in non-negligible uncertainties in the frequency of the maximum of transmission, which we estimate to ~200 MHz.

Lastly we check how the resistivity of the Al antennas affect the transmission $S$-parameter of the simulation. We check two values of resistivity, $\rho = 2.65 \times 10^{-8}$ $\Omega$m which is the value for bulk Al, and $\rho = 4 \times 10^{-8}$ $\Omega$m as an approximation for the thin sheets of Al. From Figure 11(a) we find that the amplitude for $\rho = 2.65 \times 10^{-8}$ $\Omega$m is ~1.2 times larger than for $\rho = 4 \times 10^{-8}$ $\Omega$m. It is expected that a higher resistivity of the aluminum will result in larger Ohmic losses and thus an overall lower spin wave amplitude. Other than this, there is little difference between the resistivities and the frequency position is found to be identical within the uncertainty of the simulation frequency step size. To this end we used the bulk value of resistivity for the main simulations of this study.

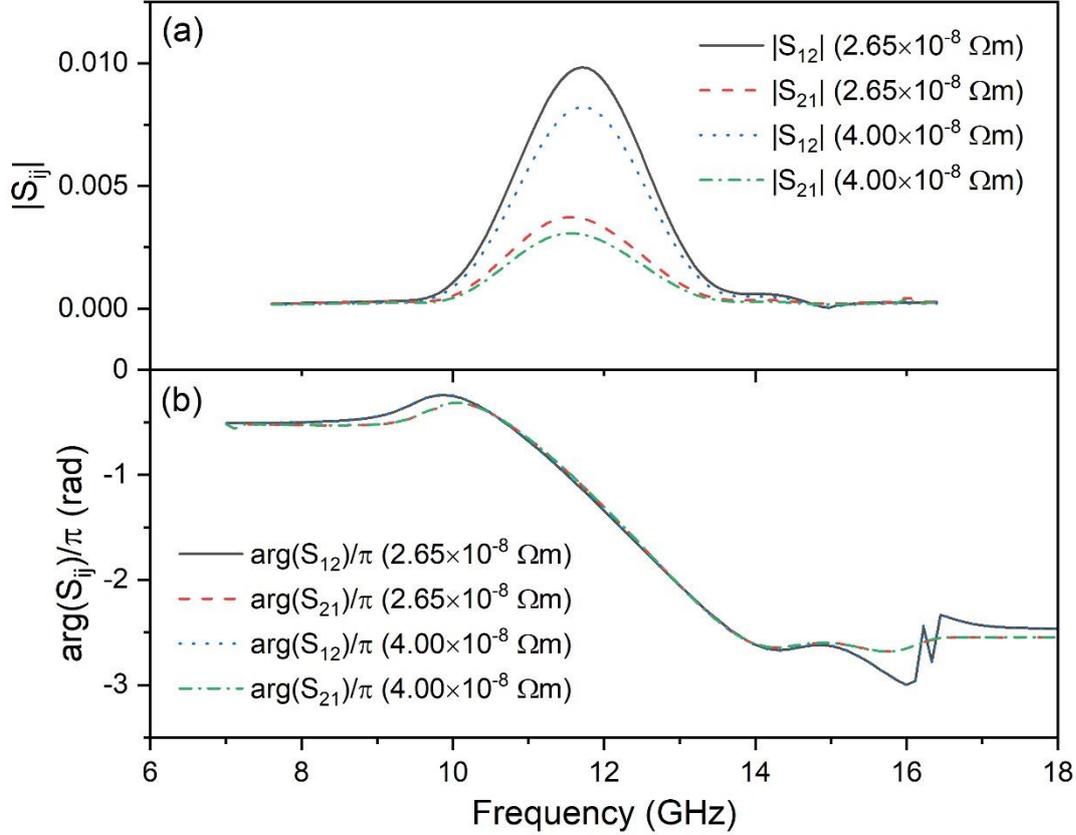

*Figure 11: (a) Amplitude and (b) phase of the transmission S-parameters for an Al antenna resistivity of 2.65× $10^{-8}$ Ω m (black solid line and red dashed line) and 4.00× $10^{-8}$ Ω m (blue dotted line and green dash dotted line).*

## 5. Conclusion

We employed the theoretical model developed in [20] to determine magnetic sample parameters of a //Ru(5)/Co(20)/Pt(5) sample by comparing the theoretical simulation results with the experimental propagating spin-wave spectroscopy measurements carried out by coplanar nano-antenna devices on the sample. In order to supplement the determination of the sample parameters additional Brillouin light scattering (BLS), ferromagnetic resonance (FMR) and magnetometry measurements were carried out. From these measurements and the theoretical simulations we obtained: $4\pi M_s$ = 17000 G, $\gamma/2\pi$ = 2.87 MHz/Oe, $A_{ex}$ = 1.78 erg/cm$^2$, and $K_{u1} = K_{u2}$ = 0 mJ/m$^2$. The frequency of the maximum of transmission for the numerical model was within 700 MHz of that for the experiment for a range of externally applied magnetic field values from ~130-1500 Oe.

It was found that the amplitude of the $S_{21}$ and $S_{12}$ transmission parameters obtained with the simpler method of de-embedding (Appendix A) were ~35 times smaller in the experiment than in the numerical simulation and this was attributed to the extra direct parasitic coupling present between the feeding lines of the antennas in the experiment. In order to confirm this claim, the feeding lines of the device were characterized and were de-embedded from the experimental S-parameters by using a more involved method of de-embedding proposed in Appendices C and D. The so de-embedded experimental $S^*$ parameters had a very good qualitative and quantitative agreement with the theoretical ones, which confirms the validity of the theoretical simulation. As a result, we claimed that the simpler de-embedding procedure (Appendix A), which does not require characterization of the feeding lines, may be employed to accurately compare the shapes and frequency positions of the experiment and simulation, and only when a comparison in amplitudes is important does one require a more involved de-embedding procedure as outlined in Appendices C and D.

## Acknowledgements

A Research Collaboration Award from the University of Western Australia (UWA) is acknowledged. C.W. acknowledges his Research Training Program stipend from UWA and a travel grant from the Franco-Australian Hubert Curien Program. The authors acknowledge the facilities, and the scientific and technical assistance of Microscopy Australia at the Centre for Microscopy, Characterisation & Analysis, The University of Western Australia, a facility funded by the University, State and Commonwealth Governments. We also acknowledge the STnano clean-room platform partly funded by the Interdisciplinary Thematic Institute QMat, as part of the ITI 2021-2028 program of the University of Strasbourg, CNRS and Inserm, IdEx Unistra (ANR 10 IDEX 0002), SFRI STRAT'US project (ANR 20 SFRI 0012) and ANR-17-EURE-0024 under the framework of the French Investments for the Future Program. We thank M. Acosta and C. Mény for their help during sputtering deposition, and R. Bernard, S. Siegwald and H. Majjad for technical support during nanofabrication work in the STnano platform.

# Appendix A: De-embedding of the experimental data

The process of de-embedding the spin-wave signal from the contribution of the feeding pads and probes is described below.

First let us consider that the vector network analyzer (VNA) measures the complex scattering parameters of the system. The scattering parameters make up an **S** matrix of the system as follows:

$$\mathbf{S}_{tot} = \begin{bmatrix} \dot{S}_{11} & \dot{S}_{12} \\ \dot{S}_{21} & \dot{S}_{22} \end{bmatrix}, \tag{A1}$$

where $S_{ii}$ refers to the reflection coefficient of port $i$ and $S_{ij}$ corresponds to the transmission coefficient from port $j$ to port $i$, where $(i, j) = (1, 2)$. In order to carry out the de-embedding we must take two measurements with the VNA, one which is on the spin-wave band and one which is off the spin-wave band. This is achieved by measuring the $S$-parameters over a given frequency range, once with a static external magnetic field applied to the sample such that the spin-wave band is within the frequency range measured by the VNA. And a second time, where the static external magnetic field is set large enough that the spin-wave band is shifted well outside the frequency band measured by the VNA. Thus we obtain two **S** matrices,

$$\mathbf{S}_{tot\,on} = \begin{bmatrix} \dot{S}_{11on}^{exp} & \dot{S}_{12on}^{exp} \\ \dot{S}_{21on}^{exp} & \dot{S}_{22on}^{exp} \end{bmatrix}, \tag{A2}$$

$$\mathbf{S}_{tot\,off} = \begin{bmatrix} \dot{S}_{11off}^{exp} & \dot{S}_{12off}^{exp} \\ \dot{S}_{21off}^{exp} & \dot{S}_{22off}^{exp} \end{bmatrix}, \tag{A3}$$

which relate to what is physically measured with the VNA.

Now let us consider the total system probed by the VNA to be three cascaded two-port networks, where the first and third network can be considered as the input and output feeding pads respectively, and the second two-port network as the spin-wave antennas with the magnetic film, as shown in Figure 11. We will consider that the feeding input and output pads are identical and thus have the same **S** matrix, and that they are the same off and on the spin-wave band,

$$\mathbf{S}_{(1)\,on} = \mathbf{S}_{(1)\,off} = \mathbf{S}_{(3)\,on} = \mathbf{S}_{(3)\,off} = \begin{bmatrix} 0 & \dot{S}_{12p} \\ \overline{\dot{S}_{12p}} & 0 \end{bmatrix}, \tag{A4}$$

where $\overline{\dot{S}_{12p}}$ is the complex conjugate of $\dot{S}_{12p}$. Here we have assumed that the feeding pads and probes have a good impedance match with the external circuit and that all of the power is transmitted through the feeding pads and probes.

The second two-port network will differ on and off the spin-wave band since it contains the contribution of the spin-waves. Thus we have,

$$\mathbf{S}_{(2)\,on} = \begin{bmatrix} \dot{S}_{11on} & \dot{S}_{12on} \\ \dot{S}_{21on} & \dot{S}_{22on} \end{bmatrix}, \tag{A5}$$

$$\mathbf{S}_{(2)\,off} = \begin{bmatrix} \dot{S}_{11off} & \dot{S}_{12off} \\ \dot{S}_{21off} & \dot{S}_{22off} \end{bmatrix}. \tag{A6}$$

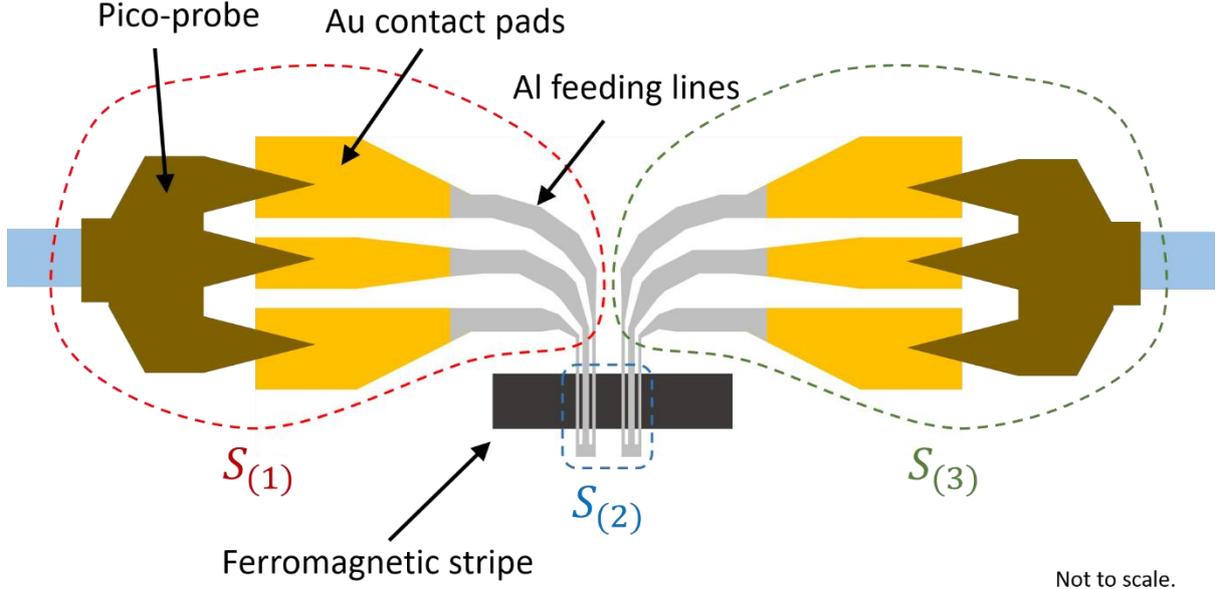

*Figure 12: Schematic of the spin-wave device showing the separation of the three cascaded two-port networks by the dashed lines. Note the schematic is not to scale and is purely intended as a guide.*

Now, in order to obtain the total $\mathbf{S}_{tot\ on}$ and $\mathbf{S}_{tot\ off}$ matrices from the individual $\mathbf{S}$ matrices of the three cascaded two-port networks we must first convert the individual $\mathbf{S}$ matrices to transmission ($\mathbf{T}$) matrices. We refer the reader to Table. VI in [31] for the conversions of $\mathbf{S}$ matrices to $\mathbf{T}$ matrices. The conversions yields,

$$\mathbf{T}_{(1,3)} = \begin{bmatrix} \dot{S}_{12p} & 0 \\ 0 & \frac{1}{\dot{S}_{12p}} \end{bmatrix}, \tag{A7}$$

$$\mathbf{T}_{(2)\ on} = \begin{bmatrix} \dot{S}_{12on} - \frac{\dot{S}_{11on}\dot{S}_{22on}}{\dot{S}_{21on}} & \frac{\dot{S}_{11on}}{\dot{S}_{21on}} \\ -\frac{\dot{S}_{22on}}{\dot{S}_{21on}} & \frac{1}{\dot{S}_{21on}} \end{bmatrix}, \tag{A8}$$

$$\mathbf{T}_{(2)\ off} = \begin{bmatrix} \dot{S}_{12off} - \frac{\dot{S}_{11off}\dot{S}_{22off}}{\dot{S}_{21off}} & \frac{\dot{S}_{11off}}{\dot{S}_{21off}} \\ -\frac{\dot{S}_{22off}}{\dot{S}_{21off}} & \frac{1}{\dot{S}_{21off}} \end{bmatrix}. \tag{A9}$$

It is well known that the total $\mathbf{T}$ matrix of a cascaded system of two-port networks is equivalent to the matrix multiplication of the individual $\mathbf{T}$ matrices of the constituent two-port networks [32]. Thus we obtain,

$$\mathbf{T}_{tot\ on} = \mathbf{T}_{(1)}\mathbf{T}_{(2)\ on}\mathbf{T}_{(3)} = \begin{bmatrix} \dot{S}_{12p} & 0 \\ 0 & \frac{1}{\dot{S}_{12p}} \end{bmatrix} \begin{bmatrix} \dot{S}_{12on} - \frac{\dot{S}_{11on}\dot{S}_{22on}}{\dot{S}_{21on}} & \frac{\dot{S}_{11on}}{\dot{S}_{21on}} \\ -\frac{\dot{S}_{22on}}{\dot{S}_{21on}} & \frac{1}{\dot{S}_{21on}} \end{bmatrix} \begin{bmatrix} \dot{S}_{12p} & 0 \\ 0 & \frac{1}{\dot{S}_{12p}} \end{bmatrix}, \tag{A10}$$

$$\mathbf{T}_{tot\ off} = \mathbf{T}_{(1)}\mathbf{T}_{(2)\ off}\mathbf{T}_{(3)} = \begin{bmatrix} \dot{S}_{12p} & 0 \\ 0 & \frac{1}{\dot{S}_{12p}} \end{bmatrix} \begin{bmatrix} \dot{S}_{12off} - \frac{\dot{S}_{11off}\dot{S}_{22off}}{\dot{S}_{21off}} & \frac{\dot{S}_{11off}}{\dot{S}_{21off}} \\ -\frac{\dot{S}_{22on}}{\dot{S}_{21off}} & \frac{1}{\dot{S}_{21off}} \end{bmatrix} \begin{bmatrix} \dot{S}_{12p} & 0 \\ 0 & \frac{1}{\dot{S}_{12p}} \end{bmatrix}. \tag{A11}$$

Following through with the matrix multiplication and simplifying yields,

$$\mathbf{T}_{tot\,on} = \begin{bmatrix} (\dot{S}_{12p})^2(\dot{S}_{12on} - \frac{\dot{S}_{11on}\dot{S}_{22on}}{\dot{S}_{21on}}) & \frac{\dot{S}_{11on}\dot{S}_{12p}}{\dot{S}_{21on}\overline{\dot{S}_{12p}}} \\ -\frac{\dot{S}_{22on}\dot{S}_{12p}}{\dot{S}_{21on}\overline{\dot{S}_{12p}}} & \frac{1}{\dot{S}_{21on}(\overline{\dot{S}_{12p}})^2} \end{bmatrix}, \quad (A12)$$

$$\mathbf{T}_{tot\,off} = \begin{bmatrix} (\dot{S}_{12p})^2(a\dot{S}_{12off} - \frac{\dot{S}_{11off}\dot{S}_{22off}}{\dot{S}_{21off}}) & \frac{\dot{S}_{11off}\dot{S}_{12p}}{\dot{S}_{21off}\overline{\dot{S}_{12p}}} \\ -\frac{\dot{S}_{22off}\dot{S}_{12p}}{\dot{S}_{21off}\overline{\dot{S}_{12p}}} & \frac{1}{\dot{S}_{21off}(\overline{\dot{S}_{12p}})^2} \end{bmatrix}. \quad (A13)$$

Now that we have obtained the total **T** matrices from the individual two-port networks, we can revert these **T** matrices back into **S** matrices using the same Table. VI in [31].

Thus, after conversion and simplification, the total **S** matrices we obtain are as follows,

$$\mathbf{S}_{tot\,on} = \begin{bmatrix} \dot{S}_{12p}\dot{S}_{11on}\overline{\dot{S}_{12p}} & \dot{S}_{12on}(\dot{S}_{12p})^2 \\ \dot{S}_{21on}(\overline{\dot{S}_{12p}})^2 & \dot{S}_{12p}\dot{S}_{22on}\overline{\dot{S}_{12p}} \end{bmatrix}, \quad (A14)$$

$$\mathbf{S}_{tot\,off} = \begin{bmatrix} \dot{S}_{12p}\dot{S}_{11off}\overline{\dot{S}_{12p}} & \dot{S}_{12off}(\dot{S}_{12p})^2 \\ \dot{S}_{21off}(\overline{\dot{S}_{12p}})^2 & \dot{S}_{12p}\dot{S}_{22off}\overline{\dot{S}_{12p}} \end{bmatrix}, \quad (A15)$$

Before proceeding with the de-embedding process we will consider that the internal two-port network described by $\mathbf{S}_{(2)\,on}$, consists of two effective parallel two-port networks, each with their own **S** matrices. The first corresponds to the spin-wave channel which accounts for the propagation of spin-waves between the two antennas, whilst the second corresponds to the direct coupling between the two antennas. Note that the spin-wave channel exists only on the spin-wave band, whilst the direct coupling channel exists both off and on the spin-wave band. The matrices follow,

$$\mathbf{S}_{sw} = \begin{bmatrix} \dot{S}_{11sw} & a\dot{S}_{12sw} \\ a\dot{S}_{21sw} & \dot{S}_{22sw} \end{bmatrix}, \quad (A16)$$

$$\mathbf{S}_{dc} = \begin{bmatrix} \dot{S}_{11dc} & a\dot{S}_{12dc} \\ a\dot{S}_{12dc} & \dot{S}_{11dc} \end{bmatrix}, \quad (A17)$$

where $\dot{S}_{iisw} = 1 - a\Delta\dot{S}_{iisw}$, $\dot{S}_{11dc} = 1 - a\Delta\dot{S}_{11dc}$, and $a$ is a smallness constant since the transmission through the spin-wave and direct coupling channels is much smaller than the reflected signal off of these channels. Note that the direct coupling channel is fully reciprocal. We will now convert these **S** matrices into admittance (**Y**) matrices since the total **Y** matrix of two parallel two-port networks is simply the sum of the individual **Y** matrices. The conversions are carried out following Table. I in [31], and yield,

$$\mathbf{Y}_{sw} = \begin{bmatrix} \frac{[(1-\dot{S}_{11sw})(1+\dot{S}_{22sw})+a^2\dot{S}_{12sw}\dot{S}_{21sw}]Y_c}{(1+\dot{S}_{11sw})(1+\dot{S}_{22sw})-a^2\dot{S}_{12sw}\dot{S}_{21sw}} & \frac{-2a\dot{S}_{12sw}Y_c}{(1+\dot{S}_{11sw})(1+\dot{S}_{22sw})-a^2\dot{S}_{12sw}\dot{S}_{21sw}} \\ \frac{-2a\dot{S}_{21sw}Y_c}{(1+\dot{S}_{11sw})(1+\dot{S}_{22sw})-a^2\dot{S}_{12sw}\dot{S}_{21sw}} & \frac{[(1+a\dot{S}_{11sw})(1-a\dot{S}_{22sw})+a^2\dot{S}_{12sw}\dot{S}_{21sw}]Y_c}{(1+\dot{S}_{11sw})(1+\dot{S}_{22sw})-a^2\dot{S}_{12sw}\dot{S}_{21sw}} \end{bmatrix}, (A18)$$

$$\mathbf{Y}_{dc} = \begin{bmatrix} \frac{[a^2\dot{S}_{12dc}^2 - \dot{S}_{11dc}^2 + 1]Y_c}{(1+\dot{S}_{11dc})^2 - a^2\dot{S}_{12dc}^2} & \frac{-2a\dot{S}_{12dc}Y_c}{(1+\dot{S}_{11dc})^2 - a^2\dot{S}_{12dc}^2} \\ \frac{-2a\dot{S}_{12dc}Y_c}{(1+\dot{S}_{11dc})^2 - a^2\dot{S}_{12dc}^2} & \frac{[a^2\dot{S}_{12dc}^2 - \dot{S}_{11dc}^2 + 1]Y_c}{(1+\dot{S}_{11dc})^2 - a^2\dot{S}_{12dc}^2} \end{bmatrix}, \quad (A19)$$

where $Y_c$ is the characteristic admittance of the feeding pads, which we consider to be 1/50 Siemens. Thus the total $\mathbf{Y}_{(2)\,on}$ matrix is simply the sum of Eq. (A18) and Eq. (A19), $\mathbf{Y}_{(2)\,on} = \mathbf{Y}_{sw} + \mathbf{Y}_{dc}$, and for the sake of brevity is not shown here. We now revert the total $\mathbf{Y}_{(2)\,on}$ matrix back to the $\mathbf{Y}_{(2)\,on}$ matrix, again using Table. I in [31]. Substitution of $\dot{S}_{iisw} = 1 - a\Delta\dot{S}_{iisw}$ and $\dot{S}_{11dc} = 1 - a\Delta\dot{S}_{11dc}$ into the resultant $\mathbf{S}_{(2)\,on}$ matrix and Taylor expanding each element of $\mathbf{S}_{(2)\,on}$ around $a$ to the first order results in,

$$\mathbf{S}_{(2)\,on} = \begin{bmatrix} 1 & a(\dot{S}_{12sw} + \dot{S}_{12dc}) \\ a(\dot{S}_{21sw} + \dot{S}_{12dc}) & 1 \end{bmatrix}. \quad (A20)$$

Similarly we can obtain the total $\mathbf{S}_{(2)\,off}$ matrix from Eq. (A20) by considering that the spin-wave channel is totally reflecting in this case, $\dot{S}_{12sw} = \dot{S}_{21sw} = 0$, and substitution of this condition into Eq. (A20) yields,

$$\mathbf{S}_{(2)\,off} = \begin{bmatrix} 1 & a\dot{S}_{12dc} \\ a\dot{S}_{12dc} & 1 \end{bmatrix}. \quad (A21)$$

Let us now substitute the elements of Eqs. (A20-A21) back into Eqs. (A14-A15) to obtain,

$$\mathbf{S}_{tot\,on} = \begin{bmatrix} \dot{S}_{12p}\overline{\dot{S}_{12p}} & (\dot{S}_{12p})^2 a(\dot{S}_{12sw} + \dot{S}_{12dc}) \\ (\overline{\dot{S}_{12p}})^2 a(\dot{S}_{21sw} + \dot{S}_{12dc}) & \dot{S}_{12p}\overline{\dot{S}_{12p}} \end{bmatrix}, \quad (A22)$$

$$\mathbf{S}_{tot\,off} = \begin{bmatrix} \dot{S}_{12p}\overline{\dot{S}_{12p}} & (\dot{S}_{12p})^2 a\dot{S}_{12dc} \\ (\overline{\dot{S}_{12p}})^2 a\dot{S}_{12dc} & \dot{S}_{12p}\overline{\dot{S}_{12p}} \end{bmatrix}. \quad (A23)$$

Recall that these total matrices, Eq. (A22-A23), are the same matrices as Eq. (A2-A3) which correspond to what is measured by the VNA, and thus we may manipulate the experimental data to perform the de-embedding. The de-embedding process is carried out in two steps. First we subtract the off spin-wave band measurement from the on spin-wave band measurement,

$$\mathbf{S}_{diff} = \mathbf{S}_{tot\,on} - \mathbf{S}_{tot\,off}, \quad (A24)$$

and then divide each element of this subtracted matrix by the corresponding element of the off spin-wave band matrix,

$$\dot{S}_{(ij)\,de-embedded} = \frac{\dot{S}_{(ij)\,diff}}{\dot{S}_{(ij)\,tot\,off}}, \quad (A25)$$

where $i, j \in \{1,2\}$ and correspond to the port indices. The resulting $\mathbf{S}_{de-embedded}$ matrix yields the final result,

$$\mathbf{S}_{de-embedded} = \begin{bmatrix} \frac{\dot{S}^{exp}_{11on} - \dot{S}^{exp}_{11off}}{\dot{S}^{exp}_{11off}} & \frac{\dot{S}^{exp}_{12on} - \dot{S}^{exp}_{12off}}{\dot{S}^{exp}_{12off}} \\ \frac{\dot{S}^{exp}_{21on} - \dot{S}^{exp}_{21off}}{\dot{S}^{exp}_{21off}} & \frac{\dot{S}^{exp}_{22on} - \dot{S}^{exp}_{22off}}{\dot{S}^{exp}_{22off}} \end{bmatrix} = \frac{1}{\dot{S}_{12dc}} \begin{bmatrix} 0 & \dot{S}_{12sw} \\ \dot{S}_{21sw} & 0 \end{bmatrix}. \quad (A26)$$

Note that the division here refers to a division on an elemental basis and not to the multiplication of the inverse matrix. Thus we end with the de-embedded experimental signal on the l.h.s in Eq. (A26) and what it physically relates to on the r.h.s of Eq. (A26).

Firstly we note that after the de-embedding we completely remove the contribution of the feeding pads from the experimental signal which is significant since the feeding pads add significant coupling which alters the shape of the total signal. Secondly we find that after de-embedding we completely remove the reflected spin wave signal from the two ports, to the first order of smallness. Thus, any spin wave signal present in the reflection parameters after de-embedding is of a 2$^{nd}$ order of smallness and thus have been ignored.

Lastly we see that the de-embedded $S_{12}$ and $S_{21}$ are the respective $S$-parameters for the spin wave channel weighted by some linear function in frequency, $\frac{1}{\dot{S}_{12dc}}$. Thus, the de-embedding process yields a very good representation of the spin-wave signal in transmission.

## Appendix B: Comparing the theoretical model with the de-embedded signal

Let us now consider the simulated spin-wave signal from the developed theoretical model. We may use the same approach for the theoretical signal as was used for the experimental signal in Appendix A. In the theoretical model, the presence of the feeding pads and probes is not accounted for and thus in this case the **S** matrices in Eq. (A4) would have transmission parameters equal to 1. However, as the full de-embedding process described in Appendix A removes the contributions of the feeding pads and probes completely from the signal we will arrive at the same final result for the theoretical signal if we perform the same de-embedding process. Thus for the theoretical signal we obtain,

$$\mathbf{S}_{de-embedded}^{theory} = \begin{bmatrix} \frac{\dot{S}_{11on}^{theory}-\dot{S}_{11off}^{theory}}{\dot{S}_{11off}^{theory}} & \frac{\dot{S}_{12on}^{theory}-\dot{S}_{12off}^{theory}}{\dot{S}_{12off}^{theory}} \\ \frac{\dot{S}_{21on}^{theory}-\dot{S}_{21off}^{theory}}{\dot{S}_{21off}^{theory}} & \frac{\dot{S}_{22on}^{theory}-\dot{S}_{22off}^{theory}}{\dot{S}_{22off}^{theory}} \end{bmatrix} = \frac{1}{\dot{S}_{12dc}} \begin{bmatrix} 0 & \dot{S}_{12sw} \\ \dot{S}_{21sw} & 0 \end{bmatrix}. \quad (B1)$$

This promotes a direct comparison between the theoretical de-embedded signal (Eq. (B1)) and the experimental de-embedded signal (Eq. (A26)).

## Appendix C: Characterization of the feeding lines

In order to characterize the feeding lines of the device, the reflection parameter of the device was measured for the case where the feeding lines were terminated with the intact CPL antenna, and a second time where the feeding lines were terminated with an open circuit. Note, in the following we consider the feeding lines as the picoprobes, Au contact pads, and Al feeding lines connecting the contact pads to the CPL antenna. These feeding lines are displayed by **S**$_{(1)}$ and **S**$_{(2)}$ in Figure 12. In Figure 13 the network diagram of the system is shown.

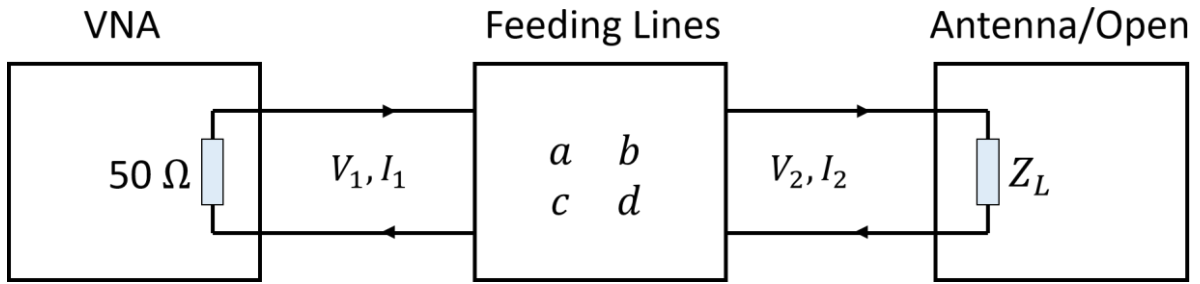

*Figure 13: Network schematic of the VNA, feeding lines, and the terminating load of the feeding line. In one case the load corresponds to the intact CPL antenna, in the second case the load corresponds to open air. The feeding lines be considered as a two port network and may be characterized by the **ABCD***

*matrix containing the a, b, c, and d parameters. $V_i$ and $I_i$ correspond to the voltage and current at port i.*

Importantly, the two ports of the feeding lines have different characteristic impedances. The characteristic impedance of the picoprobe is 50 Ohm. (We expect the characteristic impedance of the Au contact pads to also be 50 Ohm, as the cross-section of these conductors is large enough to ignore the effect of ohmic losses.) Conversely, as seen from Figure 7(b), the conductors of feeding CPL line at its junction with the antenna are quite narrow and made of Al (region near the FIB cut). Therefore, we expect ohmic resistance to dominate over in-series linear inductance in the vicinity of the feeding-line junction with the antenna and yield a characteristic impedance significantly different from 50 Ohm. For this reason we utilize the formalism of **ABCD** matrices for convenience.

If we consider the feeding lines as a two port network characterized by its **ABCD** matrix then we obtain the following equation [31],

$$\begin{bmatrix} V_1 \\ I_1 \end{bmatrix} = \begin{bmatrix} a & b \\ c & d \end{bmatrix} \begin{bmatrix} V_2 \\ -I_2 \end{bmatrix} \tag{4}$$

where, $V_i$, and $I_i$ are voltages and currents at port *i* respectively. It must stand

$$V_i = Z_L I_i \tag{5}$$

where $Z_L$ is the load impedance. In addition, the feeding lines are perfectly reciprocal which results in an additional requirement namely,

$$a \cdot d - b \cdot c = 1. \tag{6}$$

Let us now introduce incident and reflected voltages at all ports: $V_i = V_i^{inc} + V_i^{ref}$, $I_i = \frac{V_i^{inc} - V_i^{ref}}{Z_c^i}$, where $V_i^{inc}$, $V_i^{ref}$, and $Z_c^i$ are the incident voltage, reflected voltage, and characteristic impedance at port *i*, respectively. Let us consider the case of $V_1^{inc} = 1V$. Thus we get, $V_1 = 1 + S_{11}$ and $I_1 = \frac{1 - S_{11}}{Z_c^{picoprobe}}$, where $Z_c^{picoprobe}$ is the characteristic impedance of the picoprobe, which is taken to be 50 Ω. Here, $S_{11}$ is the reflection parameter from port 1 and can be measured with the VNA.

This converts Eq.(4) into,

$$\begin{bmatrix} 1 + S_{11} \\ \frac{1 - S_{11}}{Z_c^{picoprobe}} \end{bmatrix} = \begin{bmatrix} a & b \\ c & d \end{bmatrix} \begin{bmatrix} V_2 \\ -\frac{V_2}{Z_L} \end{bmatrix}. \tag{7}$$

In combination with Eq.(5), this equation has 4 unknowns: a,b,c and $V_2$. (We assume that *d* has been eliminated using Eq.(6)). If we now take three measurements of $S_{11}$ for two different values of $Z_L$, we will have 6 scalar equations for 6 unknowns – *a,b,c* and three different values of $V_2$. (Each matrix equation will generate two scalar ones.) Solving these 6 equations will complete the de-embedding process in the general case.

However, one can reduce the number of loads for which $S_{11}$ has to be measured to two and thus make this de-embedding method compatible with our approach of cutting the feeding line. To this end, at port *i*=2 we consider the incident voltage and current as those that are incident on the terminating load. Thus, $V_2^{ref} = V_2^{inc} \cdot \Gamma$, where $\Gamma$ is the reflection coefficient from the terminating load, and we can rewrite Eq.(4) as,

$$\begin{bmatrix} 1 + S_{11} \\ \frac{1 - S_{11}}{Z_c^{picoprobe}} \end{bmatrix} = \begin{bmatrix} a & b \\ c & d \end{bmatrix} V_2^{inc} \begin{bmatrix} 1 + \Gamma \\ \frac{1 + \Gamma}{Z_L} \end{bmatrix}. \tag{8}$$

Note that in the case of lumped load resistance, we must replace the characteristic impedance with the load impedance. This pertains to the right-hand side of Eq. (8).

When the feeding lines are terminated by an open circuit ($Z_L = \infty$) Eq. (8) reduces to,

$$\begin{bmatrix} 1 + S_{11}^{open} \\ \dfrac{1 - S_{11}^{open}}{Z_c^{picoprobe}} \end{bmatrix} = \begin{bmatrix} a & b \\ c & d \end{bmatrix} V_2^{inc} \begin{bmatrix} 2 \\ 0 \end{bmatrix}, \tag{9}$$

and when the feeding lines are terminated by the CPL antenna one obtains,

$$\begin{bmatrix} 1 + S_{11}^{load} \\ \dfrac{1 - S_{11}^{load}}{Z_c^{picoprobe}} \end{bmatrix} = \begin{bmatrix} a & b \\ c & d \end{bmatrix} V_2^{inc} \cdot (1 + \Gamma^{antenna}) \begin{bmatrix} 1 \\ \dfrac{1}{Z_{in}^{antenna}} \end{bmatrix}, \tag{10}$$

where $S_{11}^{open}$ and $S_{11}^{load}$ are the reflection parameters measured with the VNA, and $\Gamma^{antenna}$ and $Z_{in}^{antenna}$ are the reflection parameter and input impedance of the CPL antenna, respectively.

In order for $V_2^{inc}$ to be the same in Eqs.(8) and (9) we make an assumption, namely that the feeding lines act as a black box consisting of no sharp changes in characteristic impedance and thus act to smoothly attenuate the incident signal and smoothly transform the characteristic impedance from 50 Ohm of the picoprobe to a higher and complex-valued one at the antenna port of the feeding line. Under this assumption, which is quite realistic, the attenuation of a signal traveling in the feeding line in the forward direction does not depend on the load impedance, and the forward traveling signal will create a voltage $V_2^{inc}$ at the output port of the feeding line that will be constant for any choice of load impedance. As seen from Figure 7(b) the port represents a coplanar line with the same dimensions as the antenna. However, there is one major difference – the CPL is not backed – there is no metallic underlayer below it. Because of this peculiarity, it is useful to consider the port as a simple RC transmission line where the sheet resistance of the CPL strips is calculated for a resistivity of $2.65 \times 10^{-8}$ Ohmm, strip widths as shown in Table 1, and strip thicknesses of 100 nm. The linear capacitance is obtained from the theoretical simulation using our numerical model [20] and assuming a very large distance to the backing metal plate. The capacitance is found to be $\sim 1.96 \times 10^{-10}$ F/m.

The characteristic impedance for a particular eigen-mode of a transmission line is given by the eigen-vector of the matrix of coefficients of the Telegrapher Equations. More precisely, by the ratio of the co-efficient that has units of linear voltage to the co-efficient that has units of current. As shown in [20], in the case of a CPL, the matrix size is 5 x 5. The characteristic impedance is given by the ratio of matrix components that represent the eigen-amplitudes of the wave of voltage between the signal line and a ground line and of the wave of current in the signal line. The eigen-vector that corresponds to the CPL-like mode of the transmission line must be selected. (In addition to it, there are also slot-line-like modes and a dc-current-like mode.)

In the RC-transmission line approximation, the linear inductance of the line is neglected, which reduces the ratio of the respective components of the eigen-vector to a simple formula for the feeding line port facing the spin wave antenna:

$$Z_c^{antenna\ port} = \sqrt{\dfrac{2R_1 + R_2}{4Y_{11}}}, \tag{11}$$

where $R_1$ is the linear ohmic resistance of the signal line, $R_2$ is the linear ohmic resistance of one of the two ground lines of the CPL, and $Y_{11}$ is the linear capacitive impedance due to capacitance between the signal and one of the two ground lines, $Y_{11} = i\omega C/2$, where $\omega$ is the angular frequency, $C$ is the linear capacitance, and $i$ is the imaginary unit. The latter ($Y_{11}$) is a purely imaginary-valued quantity.

The characteristic impedance is then calculated for a range of frequencies spanning the range measured experimentally and simulated theoretically. The calculated characteristic impedance is complex-valued and is shown in Figure 14(a) by the black and red curves. We also show the characteristic impedance at the antenna port if we assume the theoretical model accurately models the amplitude of the S-parameters. In this case we solved the reverse problem of de-embedding (Appendix C) and assumed the $Z_c^{antenna\,port}$ is a free parameter and found the characteristic impedance which led to the best overlap in $|S_{21}|$ and $|S_{12}|$. This characteristic impedance is shown by the blue and green curves in Figure 14(a). The input impedance of the CPL antenna as obtained from the theoretical simulation is shown in Figure 14(b). This input impedance is almost entirely real and the average value of the real component is ~12.42 Ohm. If one calculates analytically the resistance of the CPL antenna by using the resistivity and the geometry of the antenna strips one obtains a value of ~12.56 Ohm which is consistent with the calculated input impedance of the theoretical simulation. This evidences that ohmic losses are the dominating contribution to $Z_{in}^{antenna}$.

The reflection coefficient of the antenna could then be calculated using the characteristic impedance and the input impedance of the antenna as follows,

$$\Gamma^{antenna} = \frac{Z_{in}^{antenna} - Z_c^{antenna\,port}}{Z_{in}^{antenna} + Z_c^{antenna\,port}}. \qquad (12)$$

The resultant reflection coefficient is shown in Figure 14(c).

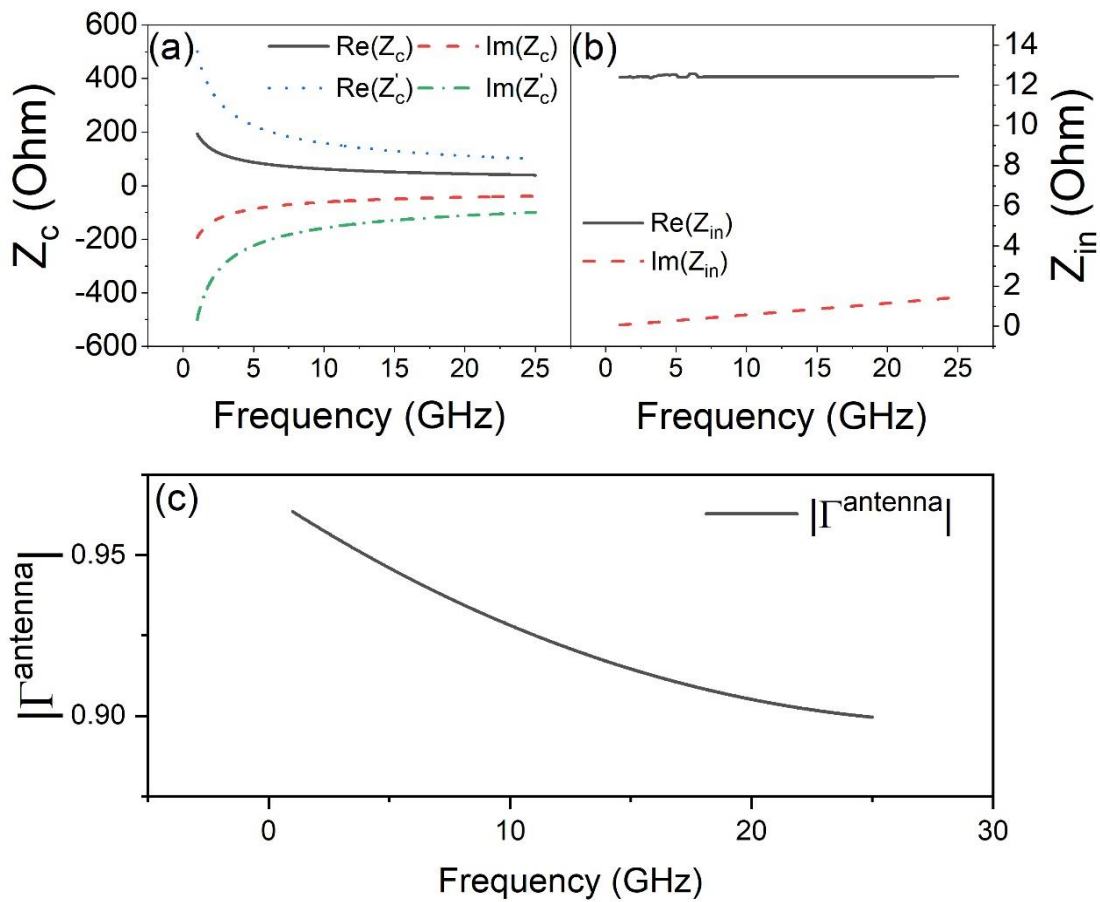

*Figure 14: (a) The real (black solid line) and imaginary (red dashed line) components of the characteristic impedance of the feeding line port facing the CPL antenna. The blue dotted and green dash-dotted line correspond to the real and imaginary components, respectively, of the characteristic*

*impedance for the feeding line port facing the CPL antenna which results in the best overlap between the de-embedded theoretical and experimental transmission $S^*$-parameters. (b) The real (black solid line) and imaginary (red dashed line) components of the input impedance of the CPL antenna. (c) Reflection coefficient of the CPL antenna.*

Once the characteristic impedance and reflection coefficient of the CPL antenna are known it is possible to solve the system of equations in Eqs. (5), (9), and (12) for the *ABCD*-parameters of the feeding line, resulting in,

$$a = -\frac{S_{11}^{open} + 1}{2\sqrt{\dfrac{-Z_c^{antenna\,port}\left(S_{11}^{load} - S_{11}^{open}\right)}{Z_c^{picoprobe}(1 - \Gamma^{antenna})}}}, \tag{13}$$

$$b = \frac{Z_c^{picoprobe}\sqrt{\dfrac{-Z_c^{antenna\,port}\left(S_{11}^{load} - S_{11}^{open}\right)}{Z_c^{picoprobe}(1 - \Gamma^{antenna})}}\left(2S_{11}^{load} - S_{11}^{open}\Gamma^{antenna} - S_{11}^{open} - \Gamma^{antenna} + 1\right)}{2\left(S_{11}^{load} - S_{11}^{open}\right)}, \tag{14}$$

$$c = -\frac{1 - S_{11}^{open}}{2Z_c^{picoprobe}\sqrt{\dfrac{-Z_c^{antenna\,port}\left(S_{11}^{load} - S_{11}^{open}\right)}{Z_c^{picoprobe}(1 - \Gamma^{antenna})}}}, \tag{15}$$

$$d = \frac{\sqrt{\dfrac{-Z_c^{antenna\,port}\left(S_{11}^{load} - S_{11}^{open}\right)}{Z_c^{picoprobe}(1 - \Gamma^{antenna})}}\left(-2S_{11}^{load} + S_{11}^{open}\Gamma^{antenna} + S_{11}^{open} - \Gamma^{antenna} + 1\right)}{2\left(S_{11}^{load} - S_{11}^{open}\right)}. \tag{16}$$

The measured $S_{11}^{open}$ and $S_{11}^{load}$ parameters are shown in Figure 15.

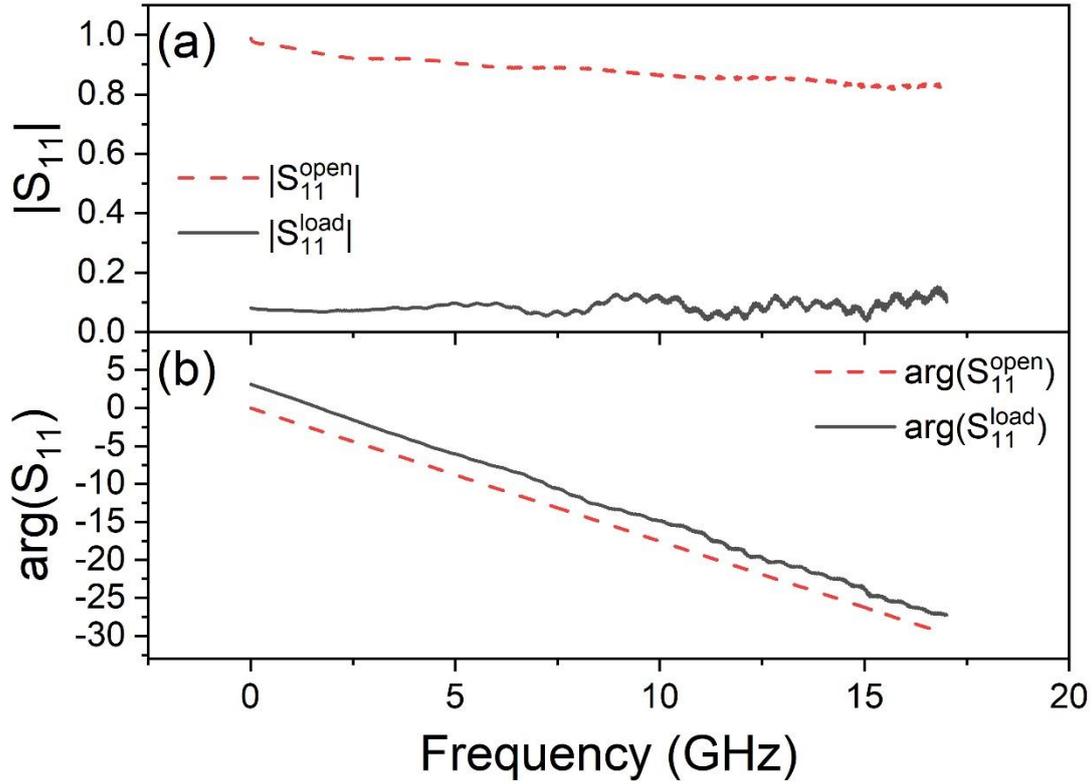

*Figure 15:(a) Amplitude and (b) phase of the real and imaginary components of the total measured reflection coefficient from the feeding line loaded by the antenna (black solid line) and when the feeding line is terminated by the open circuit (i.e. with the FIB cut, red dashed line).*

When the feeding line is terminated in an open circuit, the amplitude of the reflection coefficient is very close to one. This is expected as any signal that reaches the open circuit would perfectly reflect from it. The reflection coefficient is not perfectly one since there are expected ohmic and some small inductive and capacitive losses [33] in the system. In the case where the feeding lines are terminated by the intact antenna the amplitude of the reflection coefficient reduces substantially. This is because the CPL antenna acts as a load impedance. This large difference in the reflection coefficients then shows that the signal is almost fully damped in the CPL antennas and only a small portion returns to the port of the VNA. The phase, on the other hand, has almost the same slope for the open and loaded antenna and this is expected since the phase velocity of the electromagnetic wave in the feeding lines does not depend on the load impedance.

One can now solve for the *ABCD*-parameters using Eqs. (13)-(16). Note, that due to the significant simulation time, the reflection coefficient of the antenna was simulated at less frequency points than the experiment, and also simulated over a smaller frequency range. As a result, we opted to calculate the *ABCD*-parameters of the feeding lines for the frequencies of the simulation rather than the experiment. To do this we interpolated the $S_{11}$-parameters of the cut and loaded experimental measurements to the frequency values of the simulation. The resultant *ABCD*-parameters of the calculated feeding lines are displayed in Figure 16. Alternatively, one could instead calculate the *ABCD*-parameters of the feeding lines for the frequencies of the experimental measurements by interpolating the simulated reflection coefficient to the frequencies of the experiment. However, we find that interpolating the experimental measurements down to the simulation frequencies results in a better agreement between the de-embedded experimental data and the simulation than interpolating to the experimental frequencies. In Appendix D we show how to de-embed the characteristics of the feeding line from the total measured signals using these *ABCD*-parameters.

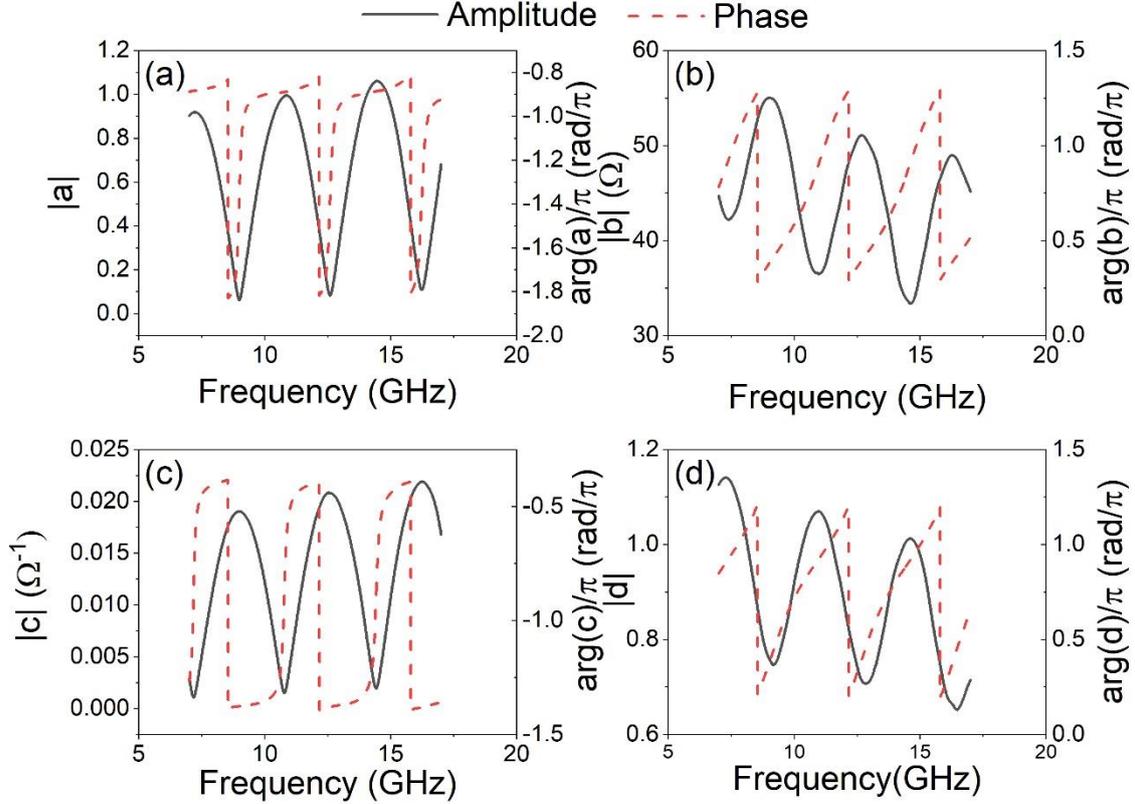

*Figure 16: Amplitude and phase of the (a) a-parameter, (b) b-parameter, (c) c-parameter and (d) d-parameter for the feeding lines. Black solid line: amplitude. Red dashed line: phase.*

# Appendix D: De-embedding of the feeding line characteristics

With the *ABCD*-parameters of the feeding lines known, it is now possible to de-embed their contribution from the total measured signal. Similar to Appendix A, we will consider the input feeding line, spin-wave delay line, and output feeding line as three cascaded two-port networks. For cascaded networks the total **ABCD** matrix of the entire system is simply the matrix multiplication of the constituent networks [31].

$$\mathbf{ABCD}_{tot} = \mathbf{ABCD}_{in} \cdot \mathbf{ABCD}_{(2)} \cdot \mathbf{ABCD}_{out}, \quad (17)$$

where $\mathbf{ABCD}_{tot}$ corresponds to the total measured system, $\mathbf{ABCD}_{in}$ corresponds to the input feeding lines which have been calculated in Appendix C, $\mathbf{ABCD}_{(2)}$ corresponds to the spin-wave delay line (2nd network in Figure 12) which contains contributions from the spin waves and the background direct inductive coupling, and $\mathbf{ABCD}_{out}$ corresponds to the output feeding lines. Note, that $\mathbf{ABCD}_{in} \neq \mathbf{ABCD}_{out}$, and $\mathbf{ABCD}_{out}$ can be obtained from $\mathbf{ABCD}_{in}$ as follows,

$$\mathbf{ABCD}_{out} = \begin{bmatrix} 1 & 0 \\ 0 & -1 \end{bmatrix} \cdot \mathbf{ABCD}_{in}^{-1} \cdot \begin{bmatrix} 1 & 0 \\ 0 & -1 \end{bmatrix} = \frac{1}{a_{in}d_{in} - b_{in}c_{in}} \begin{bmatrix} d_{in} & b_{in} \\ c_{in} & a_{in} \end{bmatrix}. \quad (18)$$

Substituting Eq. (18) into Eq. (17) and solving for $\mathbf{ABCD}_{sw}$ yields,

$$\mathbf{ABCD}_{(2)} = \begin{bmatrix} a_{(2)} & b_{(2)} \\ c_{(2)} & d_{(2)} \end{bmatrix} = \frac{1}{a_{in}d_{in} - b_{in}c_{in}} \begin{bmatrix} d_{in} & -b_{in} \\ -c_{in} & a_{in} \end{bmatrix} \cdot \begin{bmatrix} a_{tot} & b_{tot} \\ c_{tot} & d_{tot} \end{bmatrix} \cdot \begin{bmatrix} a_{in} & -b_{in} \\ -c_{in} & d_{in} \end{bmatrix} \quad (19)$$

where,

$$a_{(2)} = \frac{a_{in}(a_{tot}d_{in} - b_{in}c_{tot}) - c_{in}(b_{tot}d_{in} - b_{in}d_{tot})}{a_{in}d_{in} - b_{in}c_{in}}, \quad (20)$$

$$b_{(2)} = \frac{-b_{in}(a_{tot}d_{in} - b_{in}c_{tot}) + d_{in}(b_{tot}d_{in} - b_{in}d_{tot})}{a_{in}d_{in} - b_{in}c_{in}}, \quad (21)$$

$$c_{(2)} = \frac{-a_{in}(a_{tot}c_{in} - a_{in}c_{tot}) - c_{in}(a_{in}d_{tot} - b_{tot}c_{in})}{a_{in}d_{in} - b_{in}c_{in}}, \quad (22)$$

$$d_{(2)} = \frac{b_{in}(a_{tot}c_{in} - a_{in}c_{tot}) + d_{in}(a_{in}d_{tot} - b_{tot}c_{in})}{a_{in}d_{in} - b_{in}c_{in}}. \quad (23)$$

Before de-embedding, the experimental data was smoothed to reduce the presence of noise in the signal. Then by converting the total smoothed *S*-parameters, measured by the VNA on the spin-wave band and off the spin-wave band, into *ABCD*-parameters [31] and using the calculated *ABCD*-parameters of the feeding lines it was possible to calculate the *ABCD*-parameters of the spin-wave delay line. The amplitude and phase of these *ABCD*-parameters is shown in Figure 17 and Figure 18, respectively. At ~12 GHz, which is where the spin wave transmission is maximized, as seen from Figure 5(a), there is a clear difference between the on-band and off-band *ABCD*-parameters which can be attributed to the presence of the spin waves at this frequency. Thus converting the *S*-parameters to *ABCD*-parameters conserves the spin-wave signal. The *ABCD*-parameters were then converted back to *S*-parameters. This was done assuming the standard 50 Ohm of the characteristic impedance for the ports to which the spin wave antennas are supposed to be connected, in order to comply with the formalism we used in our numerical model. The de-embedded off-band *S*-parameters were then subtracted from the de-embedded on-band *S*-parameters in order to remove the background signal and the resultant *S*-parameters are compared to the theoretically simulated *S*-parameters in Figure 8.

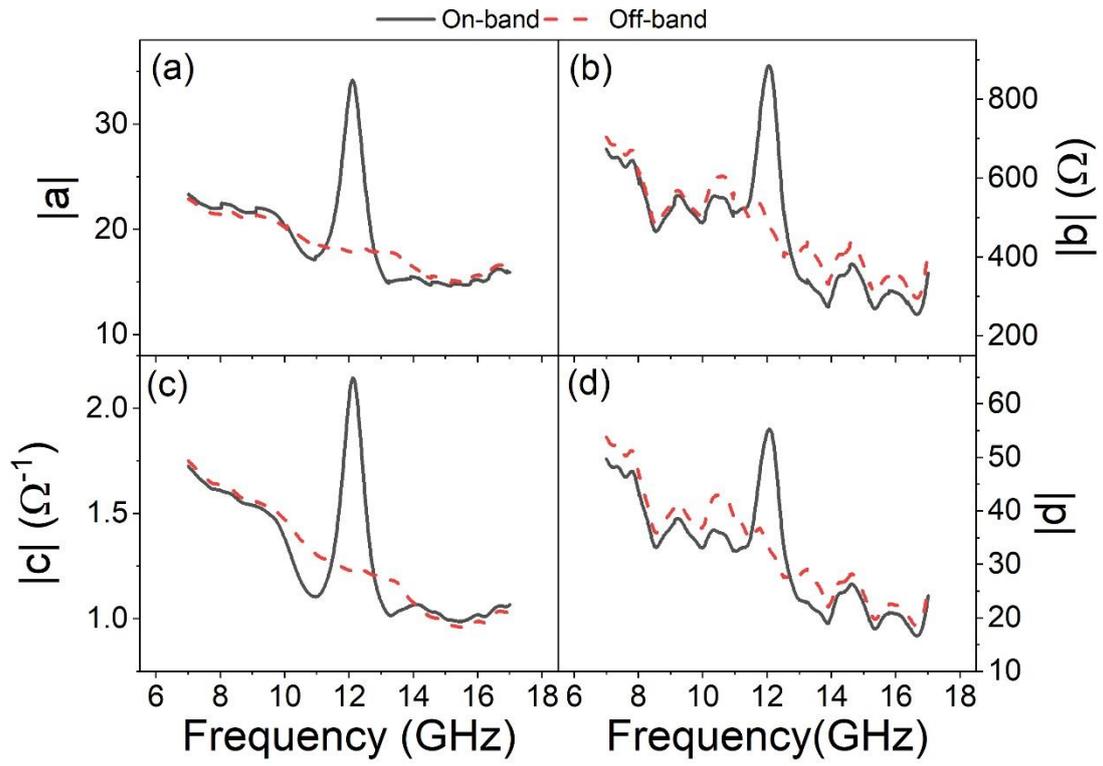

*Figure 17: Amplitude of the (a) a-parameter, (b) b-parameter, (c) c-parameter and (d) d-parameter of the de-embedded spin-wave delay line. Black solid line: on band signal. Red dashed line: off-band signal.*

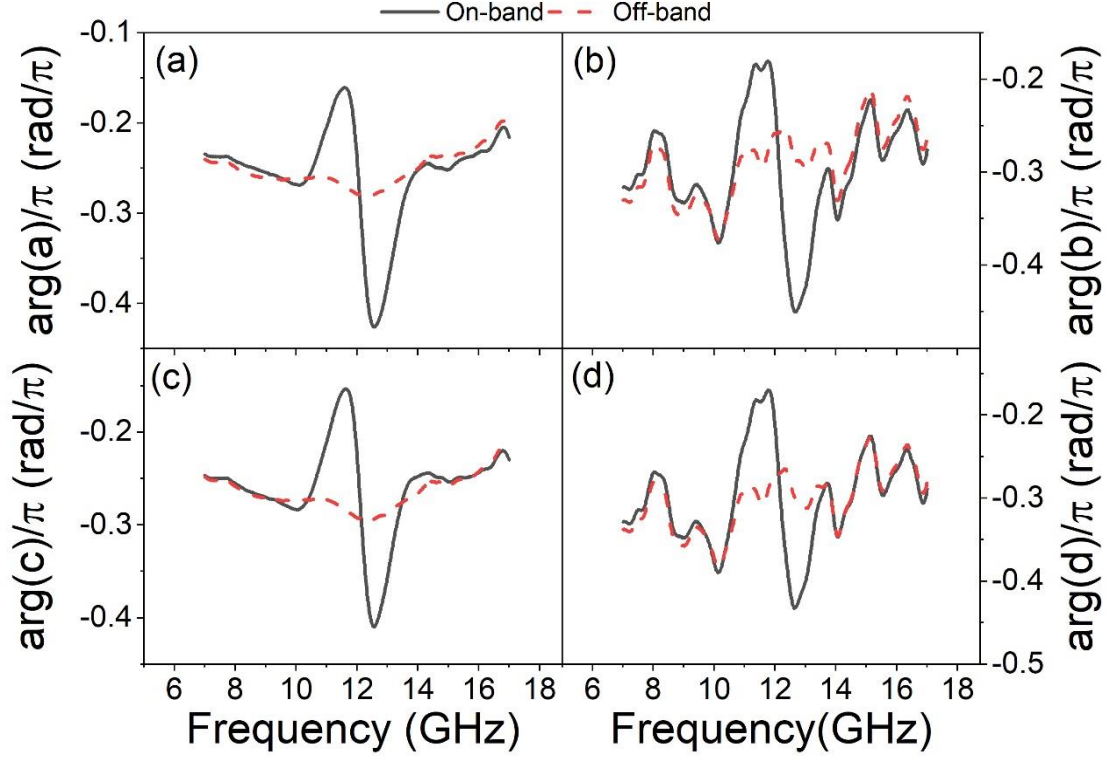

*Figure 18: Phase of the (a) a-parameter, (b) b-parameter, (c) c-parameter and (d) d-parameter of the de-embedded spin-wave delay line. Black solid line: on band signal. Red dashed line: off-band signal.*

Note that, in Figure 8 the simulated *S*-parameters are also background corrected by simulating the off-band background at a large field of 10000 Oe and subtracting the resultant *S*-parameters from the on-band *S*-parameters. The corrected parameters are denoted by $S^*$. The process of de-embedding results in a good qualitative and quantitative agreement between the transmission $S^*$-parameters of the theory and experiment, seen in Figure 8(b). This is evidenced by the very good agreement in shape and position. The amplitude of the de-embedded experimental peak and theoretically simulated one are also very similar and differ only by a factor of ~2.2, which is a significant improvement to the factor of 35 seen for the initial de-embedding procedure. The amplitude of the reflection parameters are also in good agreement, seen in Figure 8(a), with a slight difference in magnitudes of ~2. The slopes of the phases of both transmission and reflection are similar as seen in Figure 8(c) and (d). This strongly suggests that the theoretical model, underpinning the simulation, correctly models the spin wave propagation characteristics. As a result, we claim that the our initial de-embedding procedure, shown in Figure 5 and explained in Appendix A, acts as a good alternative when the characteristics of the feeding lines are not known exactly. Using this original de-embedding procedure outlined in Appendix A one obtains a good agreement in the shapes and frequency positions of the de-embedded experiment and simulated signal. The only discrepancy then remains the difference in amplitudes which can be explained as a larger background direct coupling in the experiment, and this difference in coupling can be extracted by simply rescaling the simulated data to overlap with the experiment.

It is also important to realize that the method of subtraction of the de-embedded off-band $S^*_{21}$ and $S^*_{12}$ from the respective de-embedded on-band $S^*$-parameters (Figure 8(b, d)) naturally emerges from the theory presented in Appendix A - the need to divide the anti-diagonal terms of the middle matrix of Equation (A26) by $\dot{S}^{exp}_{21\,off}$ and $\dot{S}^{exp}_{12\,off}$ arises, because this equation involves non-de-embedded *S*-parameters. If the effect of the feeding lines is excluded by the process of de-embedding $\dot{S}_{12p}$ from Equation (A22) becomes 1 (as there are no feeding lines anymore) and the anti-diagonal components of $\mathbf{S}_{diff}$ (Equation (A24)) reduce just to $\dot{S}_{21\,sw}$ and $\dot{S}_{12\,sw}$.

Conversely, the diagonal elements reduce to zeros. The latter implies that the process of subtraction of the de-embedded off-band $S_{11}^*$ and $S_{22}^*$ from the respective on-band values is not well justified physically (to the first order or smallness of $\dot{S}_{21\,sw}$ and $\dot{S}_{12\,sw}$ ).